\documentstyle[prc,aps,graphicx,floats,amsmath,amsfonts]{revtex} 

 \newcommand{\bra}{\left \langle}
 
 \newcommand{\ket}{\right \rangle}
 
 \newcommand{\Ref}[1]{(\ref{#1})}
 
 \newcommand{\Half}{\textstyle{\frac{1}{2}}}
 
 \newcommand{\I}{\imath  }
  
 \newcommand{\D}{\textstyle{\rm d}}
 
 \newcommand{\E}{\textstyle{\rm e}}
 
\begin{document}
 
 \title{\bf Determination of $\pi N$ scattering lengths from
   pionic hydrogen and pionic deuterium data} 
 \author{ A.~Deloff }
  
  \address {\normalsize Soltan Institute for Nuclear Studies,
      Hoza~69, 00-681~Warsaw, Poland }
  \maketitle
       
       \begin{abstract}
 The  $\pi$N  s-wave scattering lengths have been inferred from a joint
 analysis of the pionic hydrogen and the pionic deuterium x-ray data using a
 non-relativistic approach in which the $\pi$N interaction is simulated 
 by a short-ranged potential. This potential is assumed to be isospin
 invariant and its range, the same for isospin I=3/2 and I=1/2, is
 regarded as a free parameter. The proposed model admits an exact
 solution of the pionic hydrogen bound state problem,
 i.e. the $\pi$N scattering
 lengths can be expressed analytically in terms of the range parameter
 and the shift ($\epsilon$) and width ($\Gamma$)
 of the 1s level of the pionic hydrogen. We 
 demonstrate that for small shifts and short ranges from the exact 
 expression one retrieves the standard
 range independent Deser-Trueman formula. The $\pi$d 
 scattering length has been calculated exactly 
 by solving the  Faddeev equations
 and also by using a static approximation. It has been shown that the same
 very accurate static formula for $\pi$d scattering length can be 
 derived (i) from a set of boundary conditions; (ii) by a reduction
 of Faddeev equations; and (iii) through a summation of Feynman
 diagrams. By imposing the requirement that the $\pi$d scattering
 length, resulting from Faddeev-type calculation, be in agreement
 with pionic deuterium data, we obtain bounds on the $\pi$N
 scattering lengths.
 The dominant source of uncertainty on the deduced values
 of the $\pi$N scattering lengths are the experimental errors in the
 pionic hydrogen data.
        
 \medskip\noindent
 

 \end{abstract}
	
  \section{Introduction} 
  \label{se:one}

   The determination of low-energy pion-nucleon ($\pi$N) parameters
 has been the focus of much theoretical and experimental efforts.
 The s-wave $\pi$N scattering lengths are of particular importance
 serving as testing ground for various theoretical considerations.
 In addition to that, their isovector combination provides input 
 in the Goldberger-Miyazawa-Oehme \cite{GMO} sum rule to be used to extract
 the $\pi$NN coupling constant. In recent years major advances have been
 made in the experimental and theoretical investigation of the
 $\pi$N system. With the advent of meson factories (LAMPF, PSI and
 TRIUMF) and the corresponding influx of the new high accuracy
 $\pi$N scattering data considerable progress has been achieved in the
 $\pi$N phase shift analyses \cite{SAID,GIB98,Gashi}
 providing means to examine even such 
 subtleties as isospin symmetry breaking effects \cite{GIB98,Li,MAT97}
 Recently, the $\pi$N
 scattering experiments have been complemented by high quality pionic 
 x-ray measurements performed, both on pionic hydrogen \cite{Sigg,SCH99} 
 and on pionic deuterium \cite{HAU98}.
 The measurements of the shifts and widths in the 1s levels
 in these atomic systems, resulting from strong $\pi$N interaction,
 allows to extract directly the corresponding scattering lengths,
 i.e. $a_{\pi p}$ and $a_{\pi d}$, respectively. Therefore, the new x-ray
 data constitute an independent source of information on the low-energy
 $\pi$N scattering parameters. On the theoretical side, the physical
 quantities bearing on the low-energy $\pi$N interaction have now
 become accessible to calculations \cite{Nadia} conducted within quantum
 chromodynamics (QCD). Since QCD is known to be highly non-perturbative
 at low energies, its low-energy implementation has been based instead
 on a chiral perturbation theory in which the effective Lagrangian
 is expanded in increasing powers of derivatives in meson fields and
 quark masses. This approach in practice involves a Taylor expansion
 in the meson four-momenta and therefore it may be expected that the lower
 the energy is, the more accurate are the predictions. In this context,
 the precise knowledge of the experimental values of the low-energy
 $\pi$N scattering parameters is essential for further development of
 the theory.
 \par
 The purpose of this work is to extract the s-wave $\pi$N scattering
 lengths using exclusively the pionic hydrogen and pionic deuterium
 x-ray data.
 The key reason for proceeding along this route is that
 the low-energy regime can be thereby investigated without recourse to
 scattering data and 
 there is no danger that the low-energy parameters have been
 largely determined by the data at high energies. Our treatment
 is purely phenomenological based on an isospin invariant potential
 model and we wish to  clarify at the onset that this approach
 relinquishes any pretense of being a theory in favour of practicable
 calculational scheme.
 The investigation has two parts. In part one we take as our input
 the values of the $\pi$N scattering lengths determined previously
 from pionic hydrogen data and use them in a microscopic calculation
 of the $\pi$d scattering length. The latter has not been measured
 directly in a scattering experiment but may be 
 extracted from the pionic deuterium x-ray data by applying the
 Deser-Trueman formula \cite{Deser}.
 It is an empirical fact that the $\pi$N scattering lengths are small as
 compared with the deuteron size and it has been a common practice
 \cite{BARU}
 to use the multiple scattering expansion for calculating the $\pi$d
 scattering length. Since this series rapidly converges, what has
 been confirmed by early Faddeev calculations \cite{PETROV,AFN74,MIZ77},
 in the past with the poorly known $\pi$d scattering length  
 there was little incentive to go beyond the second order
 (for a review, cf. \cite{Judah,THO80,ERI88}). 
 At present, the experimental error on $\pi$d scattering length is
 at the level of 2\% and  the adequacy of the
 second order formula  might be 
 questionated. Strictly speaking,  a truncation
 of the multiple scattering series 
 can really only receive its justification when we
 actually quantify the magnitude of  the higher-order terms
 to establish whether they are truly negligible. 
 This question is examined in detail in this paper and 
 the $\pi$d zero-energy scattering 
 problem is solved exactly within a three-body
 formalism by introducing a zero-range
 model to simulate the $\pi$N s-wave interaction. One advantageous
 feature of this model is that it allows to obtain an analytic solution of
 the three-body problem in the static approximation. We demonstrate
 that the static solution can be obtained either by reduction of the
 Faddeev equations, or by imposing a suitable set of boundary
 conditions, or finally by performing a summation of Feynman diagrams.
 All three methods converge to the same analytic formula expressing
 the $\pi$d scattering length in terms of the $\pi$N scattering lengths.
 Static solution in coordinate space is very appealing and helps to
 develop an intuitive picture of how the individual $\pi$N 
 amplitudes contribute to build up the $\pi$d scattering length.
 By solving numerically the Faddeev equations we show that the accuracy
 of the static approximation is comparable with the present experimental
 uncertainty on $a_{\pi d}$. In order to find out what the pionic
 deuterium data can teach us about the $\pi$N scattering lengths,
 the $\pi$d scattering lengths obtained as a solution of the Faddeev
 equations is compared with experiment. It turns out that the three-body
 calculation is in agreement with experiment only when the input $\pi$N
 scattering lengths belong to a relatively small subset of values
 that are consistent with pionic hydrogen data. The $\pi$N
 scattering lengths that belong to this subset simultaneously satisfy 
 the constraints imposed by the pionic hydrogen and pionic deuterium
 data. 
 \par
 In part two of the present work 
 we introduce explicitly a range parameter 
 in order to examine the validity of the
 zero-range model. 
 To achieve this goal it is essential to devise a simple and transparent
 representation of the $\pi$N interaction in which the two-body
 scattering problem with and without Coulomb interaction admits an analytic
 solution and we show that a two-channel isospin invariant separable
 potential lends itself to that end.
 Moreover, within this representation the exact
 bound state condition appropriate for the pionic hydrogen problem
 takes also an analytic form. The latter being a single complex
 constraint, is equivalent to two real equations that can be explicitly
 solved and as  a result the $\pi$N scattering lengths are obtained
 as functions of the range parameter together with the 1s level
 shift and width in the pionic hydrogen. In particular, when the level
 shift is small as compared with the Coulomb energy and the range
 of the interaction is small in comparison with the Bohr radius,
 from the exact bound state condition we retrieve the Deser-Trueman
 formula (independent of the range parameter). 
 Regarding the range as a free parameter we are able to extend the
 zero-range model and by varying this parameter in physically reasonable
 limits we find the results to be insensitive to the value of the range.
 The uncertainty on the $\pi$N scattering length caused by the lack
 of knowledge of the range is much smaller than that resulting from
 the experimental errors on the pionic hydrogen level shift and width.
 \par
 The organization of this paper is as follows. In Sec. \ref{se:two} we
 develop  a zero-range model and review various derivations leading
 to the static solution of the $\pi$d scattering problem.
 The accuracy of the static solution is examined by comparing it with the
 solution of the Faddeev equations. We infer isoscalar and isovector
 $\pi$N scattering lengths that are consistent with both, pionic hydrogen
 and pionic deuterium data.
 In Sec. \ref{se:three} we lift the zero-range
 limitation by introducing a finite range into our formalism. 
 We present an exact treatment of the pionic hydrogen and we 
 derive Deser-Trueman formula for that particular case.
 The $\pi$d scattering length obtained from the solution of the Faddeev
 equation is compared with experiment.
 Finally, the results are summarized in Sec. \ref{se:four}. 
 
  \section{Zero-range model} 
  \label{se:two}

 The central issue we wish to address  in this section is how to
 construct a theoretical framework in which we can use the pionic
 deuterium data to gain information on the $\pi$N scattering lengths.
 The measurement of the shift and the width of the 1s level in pionic
 deuterium presents us with the value of $\pi$d scattering length
 $a_{\pi d}$. 
 The latter quantity is defined as the elastic $\pi$d scattering 
 amplitude evaluated at zero kinetic energy of the incident pion.
 This amplitude is necessarily complex because absorption reaction channels 
 are open even at the very threshold. The most important of them is the
 $\pi^{-}d\to nn$ reaction, and to a lesser extend the radiative absorption
 $\pi^{-}d\to \gamma nn$ channel. In principle, there would be also 
 the charge-exchange break-up
 channel $\pi^{-}d\to\pi^{0} nn$ that is open at threshold
 but this process is strongly suppressed by the centrifugal barrier.
 Indeed, with s-wave $\pi$N interaction there is no spin-flip possible so
 that for the two neutrons the $\mbox{}^{1}S_{0}$ state is not available,
 whereas the $\mbox{}^{3}S_{1}$ state is forbidden
 and they have to be produced in higher partial waves.
 On the whole, however, the absorptive effects are not large at
 threshold, judging from the magnitude of the imaginary part of
 the $\pi$d scattering length which empirically
 constitutes only about a quarter of
 the real part of $a_{\pi d}$. Strictly speaking, the absorptive
 processes contribute to both, the real and the imaginary part
 of $a_{\pi d}$ but in the following we are going to ignore the
 absorptive corrections to the real part of $a_{\pi d}$.
 Disregarding the absorptive processes, we shall concentrate our
 attention on a microscopic calculation of $a_{\pi d}$ and in
 order to be able to solve the ensuing three-body problem
 we introduce a potential description of the $\pi$N interaction to be
 used in the appropriate Faddeev equations.
 \par
 In order to facilitate the discussion of the Faddeev approach,
 it is instructive to take the static model as our point of
 departure. The attractive feature of the static model is that it
 is much easier to develop and to compute since the final result
 for pion-deuteron scattering length takes the form of a single analytic 
 formula that does not require off-shell information. Moreover,
 in our case the latter model also happens to be extremely good 
 approximation to the full solution of the three-body problem.
 The earliest version of a static model, due to Brueckner \cite{BRUCK},
 was based on the fixed scatterer concept and ignored
 all isospin complications. Here, we wish to make it 
 somewhat more realistic introducing as our dynamical framework 
 a set of appropriate boundary conditions, but
 on the other hand, we are prepared
 to content ourselves with a theory that has isospin invariant
 point like interactions.
 Labeling the pion as $1$ and the nucleons as $2$ and $3$,
 the boundary conditions representing the 
 zero-range $\pi$-N  interaction 
 taking place on nucleon $i$ where $i=2,3$, may be written as
 \begin{equation}
  \lim_{x_{1} \to x_{i}} 
  \overline{ |\bbox{x}_{1}-\bbox{x}_{i}| \Psi  (\bbox{x}_{1},\bbox{x}_{2}, 
   \bbox{x}_{3})}=
 (\mu/m)(b_{0}+b_{1}\,\bbox{I}\cdot \bbox{\tau}_{i})
  \lim_{x_{1} \to x_{i}} \dfrac{\D}{\D x_{1}}\;
  \overline{ |\bbox{x}_{1}-\bbox{x}_{i}| \Psi  (\bbox{x}_{1},\bbox{x}_{2}, 
  \bbox{x}_{3}) }
  \label{a1}
  \end{equation}
  where the bar denotes an average over directions
  $\bbox{x}_{1}-\bbox{x}_{i}$ what is equivalent to
   projecting out the s-wave component of the wave function $\Psi$,
  and the boundary condition \Ref{a1} is to be imposed for each of the
  two nucleons. The vectors $\bbox{I}$ and $\bbox{\tau}$ are, respectively,
  the pion and the nucleon isospin operators, whereas $b_{0}$ and $b_{1}$
  denote the isoscalar and isovector $\pi$-N scattering lengths,
  $\mu$ is the $\pi$-N reduced mass and $m$ is the pion mass.
  In the following we choose the c.m. of the two nucleons as the origin
  of  the coordinate system, i.e. we set $\bbox{x}_{2}=\Half\bbox{r}$
  and $\bbox{x}_{3}=-\Half\bbox{r}$ with $\bbox{r}$ being the
  nucleon-nucleon separation vector. The pion vector in this 
  Jacobi coordinate system will be denoted as $\bbox{\rho}$. 
  When the wave function $\Psi(\bbox{r},\bbox{\rho})$ describing
  the $\pi$NN system for the case of $\pi^{-}$ scattered off the 
  deuteron is known, the amplitude leading to the final state with
  asymptotic wave function $\Phi_{f}$ is $-\bra\Phi_{f}|V|\Psi\ket$
  where $V$ denotes the potentials that have been taken out
  in the derivation of $\Phi_{f}$. For elastic scattering 
  $\Phi_{f} ( \bbox{\rho} , \bbox{r} )=
  \exp{ (\imath \bbox{p}' \cdot \bbox{\rho} ) } \, \psi_{d}(\bbox{r}) $
  where $\bbox{p}'$ is the momentum of the
  outgoing pion, $\psi_{d}$ is the deuteron wave function and
  $V$ is the sum of the two $\pi$N potentials as asymptotically there is
  no $\pi$-deuteron interaction. Although in our formalism we never
  needed $\pi$N potentials and the $\pi$N interaction is represented by
  the boundary condition \Ref{a1}, it is in fact possible to give a
  formal expression for such potential (cf. \cite{HUANG})
  and for the operator $V$ we take
  \begin{equation}
  V\Psi(\bbox{\rho},\bbox{r})=-\frac{2\pi}{\mu}
  \left \{
  (b_{0}+b_{1}\,\bbox{I}\cdot\bbox{\tau}_{2})
  \,\delta(\bbox{\rho}-\Half\bbox{r}) \dfrac{\D}{\D\rho}
  |\bbox{\rho}-\Half\bbox{r}|+
  (b_{0}+b_{1}\,\bbox{I}\cdot\bbox{\tau}_{3})
  \,\delta(\bbox{\rho}+\Half\bbox{r}) \dfrac{\D}{\D\rho}
  |\bbox{\rho}+\Half\bbox{r}|\right\}\Psi(\bbox{\rho},\bbox{r}).
  \label{a2}
  \end{equation}
  Denoting the incident pion momentum as
   $\bbox{p}$ and 
  making use of the boundary conditions \Ref{a1} in \Ref{a2},
  the $\pi$-d elastic scattering amplitude
  $f(\bbox{p}',\bbox{p})$ takes the form
  \begin{equation}
  f(\bbox{p}^{\prime},\bbox{p})
  =\frac{\nu}{m} \int \E^{-\imath\,\bbox{p}' \cdot\bbox{\rho}}\;
   \psi_{d}^{\dagger}(\bbox{r}) \; \left \{
  \delta(\bbox{\rho}-\Half\bbox{r})
  |\bbox{\rho}-\Half\bbox{r}|+
  \delta(\bbox{\rho}+\Half\bbox{r})
  |\bbox{\rho}+\Half\bbox{r}|\right\}\Psi(\bbox{\rho},\bbox{r})
  \D^{3}\rho\;\D^{3}r, 
  \label{a3}
  \end{equation}
  where $\nu$ is $\pi$-d reduced mass. 
  Given the elastic $\pi$-d scattering  amplitude \Ref{a3}, 
  the $\pi$-d scattering length follows immediately from
  \begin{equation}
  a_{\pi d}=f(0,0).
  \label{a4}
  \end{equation}
  \par
  With the $\pi$-N interaction assumed to be isospin invariant, it will
  be convenient for us to adopt an isospin notation.
  For the initial $\pi^{-}$-d
  system, the isotopic spin wave function has the form 
  \begin{equation}
  \chi_{a}=\pi^{-}\tfrac{1}{\sqrt{2}}(p_{2}n_{3}-n_{2}p_{3}),
  \label{a5}
  \end{equation}
  where the symbols $p, n, \pi^{-}$ in \Ref{a5}
  stand for the isospin wave functions of 
  the corresponding particles. The wave function \Ref{a5} is
  antisymmetric in the nucleon labels, as appropriate for the 
  state where the isospin of the two-nucleon subsystem
  $I_{23}$ equals zero.
  As a  result of the interaction, the two nucleons
  can undergo a transition to
  a symmetric configuration corresponding to $I_{23}=1$ and we shall
  need also a function that is symmetric under two-nucleon permutation 
  \begin{equation}
  \chi_{s}= \tfrac{1}{2} \pi^{-} (p_{2}n_{3}+n_{2}p_{3})-
  \tfrac{1}{\sqrt{2}} \pi^{0}n_{2}n_{3}.
  \label{a6}
  \end{equation} 
  Since our interest here is confined to s-wave interactions, 
  no spin flip is possible and therefore
  the spin part of the wave function does not change.
  Regarding the nucleons as fixed scattering centers, 
  we may anticipate that the wave function 
  $\Psi(\bbox{\rho},\bbox{r})$ for the full system of the target
  nucleons and the meson will take the approximate form
   \begin{eqnarray}
  \Psi(\bbox{\rho},\bbox{r})=
  e^{\imath\bbox{p}\cdot\bbox{\rho}}\, 
   u_{d}(r)\, \chi_{a}
  + A(\bbox{r})  \left [  \dfrac{
    \exp{(\imath p |\bbox{\rho}-\Half\bbox{r}|)}} 
  {|\bbox{\rho}-\Half\bbox{r}|}  
  + \dfrac{
  \exp{(\imath p |\bbox{\rho}+\Half\bbox{r}|)}} 
  {|\bbox{\rho}+\Half\bbox{r}|}  \right ]   \chi_{a} + 
   \nonumber   \\
    \mbox{} + X(\bbox{r})  \left [
   \dfrac{\exp{(\imath p |\bbox{\rho}-\Half\bbox{r}|)}}  
  {|\bbox{\rho}-\Half\bbox{r}|}  
  - \dfrac{\exp{(\imath p |\bbox{\rho}+\Half\bbox{r}|)}}  
  {|\bbox{\rho}+\Half\bbox{r}|}
   \right ]  \chi_{s},
   \label{a7}
  \hspace{1cm}
   \end{eqnarray} 
  where $u_{d}$ is the spatial part of deuteron wave function that
  includes also the deuteron spin and in particular
  may contain also the D-component. The projectile enters with
  momentum $\bbox{p}$ and in the initial asymptotic region the pion and
  the target have separate wave functions (a plane wave and $u_{d}(r)$,
  respectively) and the propagation from one scattering center to
  another is described by a superposition of spherical waves.
  The hitherto unknown  amplitudes
  denoted  in \Ref{a7}, respectively,  as $A(\bbox{r})$  and $X(\bbox{r})$  
  multiplying these outgoing waves emitted by the two centers
  account for the multiple scattering phenomena. They  will
  be determined from the boundary conditions \Ref{a1}. 
  To satisfy Pauli principle the wave function \Ref{a7}
  must be antisymmetric in the two nucleon variables. 
  This implies that we have to stipulate that 
  the coefficients $A(\bbox{r})$ and $X(\bbox{r})$ are even
  under permutation of the nucleons, i.e. they
  must be invariant under the reflections $\bbox{r}\to -\bbox{r}$. 
  For zero-energy scattering  considered in this work,
  however, this is always the case because
  $A(r)$ and $X(r)$ depend then only upon the magnitude of
  $\bbox{r}$. It is worth noting that the wave function \Ref{a7}
  includes explicitly virtual charge exchange amplitude $X(r)$. 
  Since our interest here is confined to zero-energy scattering,
  in the following we take $p=0$ in \Ref{a7}. 
  Equations for the functions $A(r)$ and $X(r)$ may be obtained
  by substituting \Ref{a7} in \Ref{a1} for $i=2$ and equating the 
  coefficients multiplying the same isospin functions. With
  two different isospin functions we obtain two equations
  and this procedure  determines uniquely $A(r)$ and $X(r)$.
  Owing to the proper antisymmetrization of our wave function the  
  boundary condition for $i=3$ will be then automatically satisfied.
  The equations obtained from \Ref{a1} are
  \begin{subequations}
  \label{a8}
  \begin{eqnarray}
  A(r)&=&\tilde{b}_{0}u_{d}(r) +(\tilde{b}_{0}/r)\, A(r)+
  \sqrt{2}(\tilde{b}_{1}/r)\, X(r),
  \label{a8:a}
  \\ 
  -X(r)&= &\sqrt{2}\tilde{b}_{1}u_{d}(r) +\sqrt{2}
  (\tilde{b}_{1}/r)\,A(r)
  +(\tilde{b}_{0}+\tilde{b}_{1})/r \,X(r).  
  \label{a8:b}
  \end{eqnarray}
  \end{subequations}
  In \Ref{a8} we introduced the abbreviation $\tilde{b}_{j}=(1+m/M)b_{j}$
  where $M$ is the nucleon mass.
  The $\pi$-d scattering length is given by the overlap integral
  \begin{equation}
  a_{\pi d}=(2\nu/m)\,\int u_{d}(r)^{\dagger}A(r) \, \D^{3}r
  \label{a9}
  \end{equation}
  where $A(r)$ is the solution of \Ref{a8}   
  \begin{equation}
  A(r)= \frac{ \tilde{b}_{0}+ (\tilde{b}_{0}+\tilde{b}_{1})
  (\tilde{b}_{0}-2\tilde{b}_{1})/r }
  {1-\tilde{b}_{1}/r - (\tilde{b}_{0}+\tilde{b}_{1})
  (\tilde{b}_{0}-2\tilde{b}_{1})/r^{2} } \; u_{d}(r).
  \label{a10}
  \end{equation}
  Using \Ref{a10} in \Ref{a9} and
  expanding $A(r)$ in powers of the $\pi$N scattering lengths,
  we retrieve the well known 
  second order formula for the $\pi$-d scattering length
  (cf. \cite{ERI88})
  \begin{equation}
  a_{\pi d}^{(2)}=\frac{2\nu}{m}\left [
  \tilde{b}_{0}+(\tilde{b}_{0}^{2}-2\,\tilde{b}_{1}^{2})
  \bra \frac{1}{r} \ket \right ],
  \label{a12}
  \end{equation}
  where the expectation value is taken with respect to the
  deuteron wave function. As advertised at the beginning of this section,
  formula \Ref{a10} provides a complete solution of the problem.
  To examine the accuracy of the static formula we have to compare it
  with the exact solution of the three-body problem. The latter will be
  obtained by solving the Faddeev equations on which we now embark.
  \par
  To solve the Faddeev equations it will be convenient for us to work in
  momentum space.  Introducing the Faddeev partitions, we
  write the three-body wave function as
  \begin{equation}
  \Psi= \psi^{(1)}(\bbox{q}_{1},\bbox{k}_{1})
  + \psi^{(2)}(\bbox{q}_{2},\bbox{k}_{2})
  + \psi^{(3)}(\bbox{q}_{3},\bbox{k}_{3}),
  \label{a15}
  \end{equation}
  where $\bbox{q}_{1}$ denotes the relative momentum of the (23) pair
  whereas $\bbox{k}_{1}$ is the c.m. momentum of particle  1 
  and cyclic permutations are implied. To obtain Faddeev equations for
  the amplitudes, the different partitions are written as (cf.
  \cite{PETROV})
  \begin{subequations}
  \label{a16}
  \begin{eqnarray}
  \psi^{(1)}(\bbox{q},\bbox{k})&=&(2\pi)^{3}\phi(\bbox{q})
  \delta(\bbox{k}-\bbox{p}) \chi_{a}
  +[F(\bbox{q},\bbox{k}) \chi_{a}+G(\bbox{q},\bbox{k})\chi_{s}]
  /(E-q^{2}/M-k^{2}/2\nu);
  \label{a16:a}
   \\
  \psi^{(2)}(\bbox{q},\bbox{k})&=&
  [A(-\bbox{q},\bbox{k})\chi_{a} - X(-\bbox{q},\bbox{k})\chi_{s}] 
  /(E-q^{2}/2\mu-k^{2}/2\nu_{N});
  \label{a16:b}
   \\
  \psi^{(3)}(\bbox{q},\bbox{k})&=&
  [A(\bbox{q},\bbox{k})\chi_{a} + X(\bbox{q},\bbox{k})\chi_{s}] 
  /(E-q^{2}/2\mu-k^{2}/2\nu_{N});
  \label{a16:c}
  \end{eqnarray}
  \end{subequations}
  where $\nu_{N}$ is the reduced mass of the nucleon and that of
  the $\pi$N pair,  $E$ is the c.m. three particle kinetic
  energy and $\phi(\bbox{q})$ is the
  deuteron wave function in the momentum space.
  In \Ref{a16} we have introduced four scattering amplitudes 
  $F(\bbox{q},\bbox{k}), G(\bbox{q},\bbox{k}), A(\bbox{q},\bbox{k})$ and 
  $X(\bbox{q},\bbox{k})$. However, the amplitude $G(\bbox{q},\bbox{k})$
  to be  non-zero  requires at least p-wave  NN interaction and 
  therefore will be excluded from our considerations, while   
  the three remaining amplitudes
  will be determined from the Faddeev equations.
  It is evident from \Ref{a16} that under the $P_{23}$ permutation
  $\psi^{(1)} \to -\psi^{(1)}$ and 
  $\psi^{(2)} \leftrightarrow -\psi^{(3)}$, so that
  the total wave function is, 
  as required,  antisymmetric in the nucleon labels.
  Assuming exact isospin conservation, we can write the Faddeev
  equations 
  \begin{subequations}
  \label{a17}
  \begin{eqnarray}
  \lefteqn{}
  F( \bbox{q}, \bbox{k} ) = 
  \int \frac{\D^{3}\,k^{\prime}}{(2\pi)^{3}}
  \,\frac{
   \bra \bbox{q}| t(E-k^{2}/2\nu) | \Half\bbox{k}+\bbox{k}^{\prime}\ket+ 
   \bra \bbox{q}| t(E-k^{2}/2\nu) | -\Half\bbox{k}-\bbox{k}^{\prime}\ket} 
   {E-(\bbox{k}+\mu\bbox{k}^{\prime}/M)^{2}/2\mu-k^{\prime 2}/2\nu_{N}}
  \; A(\bbox{k}+\bbox{k}^{\prime}\frac{\mu}{M},\bbox{k}^{\prime});
  \label{a17:a}
  \end{eqnarray}
  \begin{eqnarray}
  \lefteqn{}
  &&
  A(\bbox{q},\bbox{k})=
  \bra \bbox{q}
  \left | t_{0}(E-\frac{k^{2}}{2\nu_{N}}) \right |\frac{\mu}{M} \bbox{k}
  +\bbox{p} \ket \,  \phi(\bbox{k}+\Half \bbox{p})+
   \nonumber
    \\
    && \hspace{1cm}
  + \int  \frac{\D^{3}\,k^{\prime}}{ (2\pi)^{3} }
  \,\frac{\bra \bbox{q}| t_{0}(E-k^{2}/2\nu_{N}) |\mu
                          \bbox{k}/M+\bbox{k}^{\prime}\ket }
   { E-(\bbox{k}+\Half\bbox{k}^{\prime})^{2}/M-k^{\prime 2}/2\nu } 
  \; F(-\bbox{k}-\Half\bbox{k^{\prime}},\bbox{k^{\prime}})+
 \nonumber
   \\
   && \hspace{1cm}
  + \int \frac{\D^{3}\,k^{\prime}}{ (2\pi)^{3} }
  \,\frac{  \bra \bbox{q}| t_{0}(E-k^{2}/2\nu_{N})
                             |-\mu\bbox{k}/m-\bbox{k^{\prime}}\ket} 
  { E-(\bbox{k}+\mu\bbox{k^{\prime}}/m)^{2}/2\mu-k^{\prime 2}/2\nu_{N} } 
 \; A(-\bbox{k}-\frac{\mu}{m}\bbox{k^{\prime}},\bbox{k^{\prime}})+
 \nonumber
   \\
   && \hspace{1cm}
 +\sqrt{2} \int \frac{\D^{3}\,k^{\prime}}{(2\pi)^{3}}
 \,\frac{ \bra \bbox{q}| t_{1}(E-k^{2}/2\nu_{N})
                           |-\mu\bbox{k}/m-\bbox{k^{\prime}}\ket } 
  { E-(\bbox{k}+\mu\bbox{k^{\prime}}/m)^{2}/2\mu-k^{\prime 2}/2\nu_{N} } 
  \;X(-\bbox{k}-\frac{\mu}{m}\bbox{k^{\prime}},\bbox{k^{\prime}});
  \label{a17:b}
  \end{eqnarray}
  \begin{eqnarray}
  \lefteqn{}
  &&
  -X(\bbox{k},\bbox{k})=\sqrt{2}
  \bra \bbox{k}
  \left |t_{1}(E-\frac{k^{2}}{2\nu_{N}}) \right |\frac{\mu}{M}\bbox{k}
  +\bbox{p} \ket \phi(\bbox{k}+\Half\bbox{p})+
  \nonumber 
  \\
   && 
  +\sqrt{2} \int \frac{\D^{3}\,k^{\prime}}{(2\pi)^{3}}
  \,\frac{\bra \bbox{k}| t_{1}(E-k^{2}/2\nu_{N})
                 |\mu\bbox{k}/M+\bbox{k^{\prime}}\ket }
   {E-(\bbox{k}+\Half\bbox{k^{\prime}})^{2}/M-k^{\prime 2}/2\nu} 
  \; F(-\bbox{k}-\Half\bbox{k^{\prime}},\bbox{k^{\prime}})+
  \nonumber
  \\
   && 
  + \int \frac{\D^{3}\,k^{\prime}}{(2\pi)^{3}}
  \,\frac{\bra \bbox{k}| [t_{0}(E-k^{2}/2\nu_{N})  
  - t_{1}(E-k^{2}/2\nu_{N})] |-\mu\bbox{k}/m-\bbox{k^{\prime}}\ket} 
  { E-(\bbox{k}+\mu\bbox{k^{\prime}}/m)^{2}/2\mu-k^{\prime 2}/2\nu_{N} } 
  \; X(-\bbox{k}-\frac{\mu}{m}\bbox{k^{\prime}},\bbox{k^{\prime}})+
  \nonumber
  \\
   && 
  +\sqrt{2} \int \frac{\D^{3}\,k^{\prime}}{(2\pi)^{3}}
  \,\frac{ \bra \bbox{k}| t_{1}(E-k^{2}/2\nu_{N})
                      |-\mu\bbox{k}/m-\bbox{k^{\prime}}\ket  } 
  { E-(\bbox{k}+\mu\bbox{k^{\prime}}/m)^{2}/2\mu-k^{\prime 2}/2\nu_{N} } 
  \; A(-\bbox{k}-\frac{\mu}{m}\bbox{k^{\prime}},\bbox{k^{\prime}});
  \label{a17:c}
  \end{eqnarray}
 \end{subequations}
  where in \Ref{a17} $\bra \bbox{q}'|t(E)| \bbox{q} \ket $
  is the NN scattering t-matrix for zero isospin and  
   $\bra \bbox{q}'|t_{j}(E)| \bbox{q} \ket $ are, respectively,
 the isoscalar $(j=0)$ and isovector $(j=1)$ $\pi$N scattering t-matrices.
 The elastic scattering amplitude is given by the expression
 \begin{equation}
 f(\bbox{p}',\bbox{p})=\lim_{p'\to p}\frac{{p'}^{2}-p^{2}}{4\pi}
 \int \phi(\bbox{q})^{\dagger}\frac{F(\bbox{q},\bbox{p'})}
 {E-q^{2}/M-{p'}^{2}/2\nu}\,\frac{\D^{3}q}{(2\pi)^{3}}
 \label{a20}
 \end{equation}
 and the scattering length is obtained from \Ref{a4}. We can use
 \Ref{a17:a} to eliminate $F(\bbox{q},\bbox{k})$ in \Ref{a20} in favour of 
 the amplitude $A(\bbox{q},\bbox{k})$. In the NN scattering matrices 
 occurring in \Ref{a17}, as a result of the limiting
 procedure, only the deuteron pole contributes and   
 scattering length is given as an overlap integral
 \begin{equation}
 a_{\pi d}= -\frac{\nu}{\pi}  \int \phi(k)^{\dagger}
 A(\bbox{k}\,\frac{\mu}{M}, \bbox{k})\,\frac{\D^{3}k}{(2\pi)^{3}}.
 \label{a21}
 \end{equation}
 The above formula is analogous to \Ref{a9}, and, in fact, the static
 approximation results \Ref{a9}-\Ref{a10} could have been derived from
 the Faddeev formalism. 
 In order to demonstrate that \Ref{a9}-\Ref{a10}  
 follow from \Ref{a17}
 we note that when the nucleons are static
 they are not supposed to scatter ($t \to 0 $) and the amplitude
 $F(\bbox{q},\bbox{k})$ drops out in \Ref{a17:b} and \Ref{a17:c}
 so that we are left with only two coupled integral equations.  
 When the underlying forces are of zero range,  
 the off-shell $\pi$N scattering amplitudes can be simplified, 
 and in that case
 $$
 \bra \bbox{q}'|t_{j}(E)|\bbox{q} \ket=
     -(2\pi/\mu)\,b_{j}/(1+\kappa b_{j}),\qquad j=0,1;
 $$
 where $ \kappa^{2}=2\mu B $ and $B$ is the binding energy
 of the deuteron. The important consequence of the zero-range
 assumption, apparent from the above formula, is that 
 the t-matrices become
 independent upon the off-shell momenta. 
 Therefore, the amplitudes $A(\bbox{q},\bbox{k})$ and
 $X(\bbox{q},\bbox{k})$ will be functions of one variable only and it
 will be convenient for us to introduce a notation that emphasizes that
 fact, setting $A(\bbox{q},\bbox{k})=-(m/2\pi){\cal A}(k)$ and
 $X(\bbox{q},\bbox{k})=-(m/2\pi){\cal X}(k)$, where ${\cal A}(k)$ and ${\cal
 X}(k)$ are two, hitherto unknown amplitudes. With static nucleons,  
 the energy denominators in
 \Ref{a17:b} and \Ref{a17:c} become all equal to
 $-B-(\bbox{k}'+\bbox{k})^{2}/2m$ 
 and  we end up with the following set of integral equations
 for the amplitudes ${\cal A}(k)$ and ${\cal}X(k)$:
 \begin{subequations}
 \label{a23}
 \begin{equation}
 {\cal A}(k)= \hat{b}_{0} \phi(k) + 
 4\pi \;\hat{b}_{0} 
 \int \frac{\D^{3} k'}
 {(2\pi)^{3}} \frac{{\cal A}(k')}{\kappa^{2}+(\bbox{k}'+\bbox{k})^{2}}
 +\sqrt{2} \; 4\pi \hat{b}_{1}
 \int \frac{\D^{3} k'}
 {(2\pi)^{3}} \frac{{\cal X}(k')}{\kappa^{2}+(\bbox{k}'+\bbox{k})^{2}};
  \label{a23:a}
 \end{equation}
 \begin{eqnarray}
 -{\cal X}(k)=  \sqrt{2}\hat{b}_{1} \phi(k)  
 +4\pi 
 ( \hat{b}_{0}-\hat{b}_{1} )
   &&  \int             \frac{\D^{3} k'} {(2\pi)^{3}} 
 \frac{{\cal X}(k')}{\kappa^{2}+(\bbox{k}'+\bbox{k})^{2}}+
 + \sqrt{2}\;4\pi \hat{b}_{1}
  \int   \frac{\D^{3} k'} {(2\pi)^{3}}  
 \frac{{\cal A}(k')}{\kappa^{2}+(\bbox{k}'+\bbox{k})^{2}},
 \label{a23:b}
 \end{eqnarray}
 \end{subequations}
 where 
 \begin{equation}
 \hat{b}_{j}=b_{j}\,(1+m/M)/(1+\kappa b_{j});\qquad j=0,1.
 \label{a24}
 \end{equation}
 
 The above set of integral equations  can be immediately
 solved by introducing the Fourier transform
 \begin{equation}
 A(r)=
 \int e^{\imath \bbox{k}\bbox{r}}  {\cal A}(k) \D^{3}k 
 \label{a25}
 \end{equation}
 together with a similar relationship for ${\cal X}(k)$ and $\phi (k)$
 and using the well known formula
 $$
 \frac{4\pi}{\kappa^{2}+(\bbox{k}+\bbox{k}')^{2}}=\int 
 e^{-\imath (\bbox{k}+\bbox{k}')\bbox{r} }\;
 \frac{e^{ -\kappa r}}{r} \D^{3} r.
 $$ 
 In order to solve  \Ref{a23}  we multiply the latter
 equations by $e^{\imath \bbox{k}\bbox{r}}$ and subsequently integrate
 them over  $\bbox{k}$. 
 As a result, we obtain a set of two algebraic equations
 for $A(r)$ and $X(r)$ that differ from
  \Ref{a8} only by $\exp{(-\kappa r)}/r$ replacing $1/r$
  and $\hat{b}_{j}$ 
  replacing $\tilde{b}_{j}$.
  Since \Ref{a21} goes over into \Ref{a9}, we are led
  to the extension of the static formula \Ref{a10}
  \begin{equation}
  A(r)= \frac{ \hat{b}_{0}+ (\hat{b}_{0}+\hat{b}_{1})
  (\hat{b}_{0}-2\hat{b}_{1})e^{-\kappa r} /r }
  {1-\hat{b}_{1} e^{-\kappa r}/r - (\hat{b}_{0}+\hat{b}_{1})
  (\hat{b}_{0}-2\hat{b}_{1})e^{-2\kappa r}/r^{2} } \; u_{d}(r).
  \label{a26}
  \end{equation}
   This formula is to be used in \Ref{a9} but now accounts for
  the binding energy correction. 
  \par
   Concluding our discussion of the static model  
   we wish to recall that a closed form expression for $\pi$d
   scattering length has been also obtained by effecting
   an explicit summation of Feynman diagrams and the
   most complete treatment can be found in Ref. \cite{Victor}.
   The ultimate static formula for $a_{\pi d}$,
   that takes into account isospin degree
   of freedom, given in \cite{Victor}
   is rather complicated and at first sight
   appears to be different from \Ref{a26}. However, a closer inspection
   reveals that the authors of Ref. \cite{Victor} 
   apparently did not realize that  their  
   fractional formula for $a_{\pi d}$ could have been significantly 
   simplified because a common factor equal
   $$
    1+\tilde{b}_{1}e^{-\kappa r}/r-(\tilde{b}_{0}+\tilde{b}_{1})
    (\tilde{b}_{0}-2\tilde{b}_{1})e^{-2\kappa r}/r^{2}
   $$
   may be pulled out both, from the numerator, as well as from the 
   denominator and eventually drops out. Indeed,
   when the redundant factor has been cancelled, the resulting expression
   is identical with \Ref{a26}. 
   Therefore, when binding corrections are disregarded,  this
   approach reproduces the static model result \Ref{a10} and it is
   reassuring that in this case all three methods give the same answer.
  \par
   To improve upon the static model one needs a numerical solution of
   the Faddeev equations and  in the following, 
   similarly as in the previous calculations \cite{PETROV,AFN74,MIZ77},
   in order to reduce the computational effort,
   all the pairwise interactions invoked will be represented
   by rank-one separable potentials. The $\pi$N s-wave interaction
   is taken in the form of a standard Yamaguchi potential with the same
   form factor in both isospin states. Since the inverse range parameter
   $\beta$  that enters that form factor is not known, similarly as 
   before, we consider
   the zero-range limit, i.e. $\beta \to \infty$. 
   For an assigned value of $\beta$, the strength parameter of the
   potential may be eliminated in favour of the scattering length and 
   the appropriate s-wave t-matrices, are 
   \begin{equation}
   \bra k|t_{j}(E)|k' \ket=-\frac{2\pi}{\mu}
   \frac{1}{1+k^{2}/\beta^{2}}\;
   \frac {b_{j}}
   {1-\imath p b_{j} \, (1-2\imath p/\beta)(1-\imath p/\beta)^{-2}}   
   \; \frac{1}{1+{k'}^{2}/\beta^{2}}
   \label{a27}
   \end{equation}
   where $p=\sqrt{2\mu E}$ and $j=0,1$ and it is evident from \Ref{a27}
   that the zero-range limit can be effected.
   When the nucleon motion is taken into account, the p-wave $\pi$N
   interaction gives contribution to the $\pi$d scattering amplitude
   even at threshold. Therefore,
   in addition to s-wave, we are going to include
   also the p-wave interaction, limiting ourselves only to the P33 wave
   as in that case both the strength and the statistical weight are dominant
   rendering the remaining p-waves  negligible. 
   The corresponding p-wave form factor of the form
   $$
   g_{\Delta}(k)=k/(k^{2}+\beta_{\Delta}^{2})
   $$
   has been adopted from \cite{AFN74} 
   with $\beta_{\Delta}=5.33\, fm^{-1}$ 
   where the depth of the separable potential can be adjusted to the
   experimentally known value of the P33 scattering volume
   taken to be $0.64\,fm^{3}$. It is well known that with the above
   form, the shape of the delta resonance cannot be well reproduced 
   but this is less important here, the essential thing is to have the P33
   amplitude at threshold correctly reproduced.
   Besides, the p-wave constitutes
   only a small correction and using a more complicated model
   does not seem to be currently justified.
   For the NN interaction we use two separable models: a simple 
   Hulthen-Yamaguchi potential with inverse range parameter equal
   $\beta_{N}=6.01162\, \sqrt{MB} $ whose strength is fixed by the 
   deuteron binding energy, and the PEST potential
   constructed in Ref. \cite{PEST} 
   with a more sophisticated form factor of the form
   \begin{equation}
   g(k)=\sum_{i=1}^{6}\frac{C_{i}}{k^{2}+\beta_{i}^{2}},
   \label{a28}
   \end{equation}
   where the parameters $C_{i}$ and $\beta_{i}$ have been tabulated in
   Ref. \cite{PEST}.
   This potential has been devised in such a way that the
   corresponding NN half-off-shell T-matrix has the same behaviour as
   that of the Paris potential \cite{Paris}. This separable replica
   of the Paris potential takes into account the 
   short range repulsion that is absent in the Yamaguchi potential
   yet retaining the simplicity of the latter.
  \par
  Using standard partial wave projections  
  the Faddeev equations \Ref{a17} can be reduced to
  a system of four coupled inhomogeneous integral equations 
  in a single variable that
  are amenable for numerical treatment.
  In the actual practice, in order to cross-check our
  numerical procedures, we used two independent methods of solving
  these equations. The direct method introduces an integration mesh
  what allows us to replace integrals by sums so that the integral
  equations take the form of a system of linear algebraic equations
  easily solvable by standard methods. The second method solves 
  the system of integral equations by successive iterations. 
  The iterative procedure 
  is equivalent to a power expansion in $\pi$N scattering lengths what  
  allows tracing down the contribution from the different orders. 
  Since the scattering lengths are rather small, as compared with
  the deuteron size,
  the iterative sequence proves to be rapidly convergent. 
  \par
      The experimental $\pi$-d scattering length 
  has been extracted  form the 1s level
  shift in pionic deuterium by using the Deser-Trueman \cite{Deser} formula.
  Therefore, the extracted quantity is in fact the Coulomb corrected
  scattering length, denoted hereafter as $a_{\pi d}^{c}$, 
  and before confronting the calculated pion-deuteron scattering length
  with experiment one needs the experimental value of $a_{\pi d}$, i.e. 
  the purely nuclear scattering length.
  Of course, Coulomb correction could be anticipated to be very small
  but since the experimental errors are also small, it is of interest
  to give some quantitative estimate of the Coulomb correction.
  In principle, for calculating the latter one needs to know
  the pion-deuteron nuclear potential responsible for the level shift.
  This potential is not known but 
  with the zero-range potential simulating the 
  $\pi$N interaction  in the first approximation it is reasonable to 
  expect that the effective potential is proportional to the nuclear density,
  so that the shape of the nuclear potential is given by the square of
  the deuteron wave function $u_{d}(2r)^{2}$.
  Still, the depth is not known, but on the nuclear scale
  this potential must be rather weak because the experimental value of 
  $a_{\pi d}^{c}$ is quite small.
  Therefore, to quantify the value of  
  the ratio of $a_{\pi d}^{c}/a_{\pi d}$ it is sufficient to
  keep only the first order terms in the nuclear potential.
  Since in this case the potential depth drops out,   
  we are led to the formula
  \begin{equation}
  \frac{a_{\pi d}^{c}}{a_{\pi d}}=
  \frac{\int_{0}^{\infty} u_{d}(2r)^{2}\,\phi_{0}(0,r)^{2}\, \D r}
       {\int_{0}^{\infty} u_{d}(2r)^{2}\,r^{2} \, \D r},
  \label{a13}
  \end{equation}
  where $\phi_{\ell}(k,r)$ denotes the regular Coulomb wave function
  that for zero-momentum (k=0) and zero 
  orbital momentum ($\ell$=0), simplifies to the form 
  \begin{equation}
  \phi_{0}(0,r) = r\;J_{1}(2\sqrt{2\nu \alpha r})/\sqrt{2\nu\alpha r} 
  \label{a14}
  \end{equation}
  where   $\alpha$ is the fine structure
  constant and $J_{1}(x)$ denotes the Bessel function. 
  Expanding \Ref{a13} in powers of $\alpha$, 
  we obtain quite adequate first 
  order formula $a_{\pi d}^{c}/a_{\pi d}=1-\alpha\nu \bra r \ket$
  where the expectation value is with respect to the deuteron wave
  function.
  We have checked that for a variety of deuteron wave functions 
  the calculated ratio \Ref{a13} has ben very stable and its 
  numerical value is 0.985.  
  Using this number together with the experimental value of      
  $a_{\pi d}^{c}$
  $$
  a_{\pi d}^{c}= [-(2.61 \pm 0.05)+\imath\,(0.63 \pm
                                0.07)]\times 10^{-2}/m_{\pi}   
  $$ 
  taken from  Ref. \cite{HAU98} where $m_{\pi}$ is the mass of
  the charged pion, we deduce the  value of
  the purely nuclear $\pi$d scattering length
  \begin{equation}  
  a_{\pi d}=(-2.65\pm 0.05)\times 10^{-2}/m_{\pi},
  \label{a29}
  \end{equation}
  and hereafter the above number 
  will be referred to as the ''experimental'' $\pi$d scattering length
  in which all absorptive effects have been neglected.
  \par
  Adopting the zero-range model of the  $\pi$N interaction, 
  for calculating the $\pi$d scattering length one needs as input just   
  the isoscalar and the isovector $\pi$N  scattering  scattering lengths.
  The values of $b_{0}$ and $b_{1}$   that have been extracted from 
  the pionic hydrogen data in Ref. \cite{SCH99}, are 
  \begin{equation}
          b_{0}=-(0.22 \pm 0.43)\times 10^{-2}/m_{\pi};\qquad
          b_{1}=-(9.05 \pm 0.42)\times 10^{-2}/m_{\pi},
  \label{a30}	  
  \end{equation}
  where the quoted uncertainty comprises the experimental errors
  together with the uncertainty introduced by applying a specific procedure
  that allows to deduce $b_{0}$ and $b_{1}$ from the 
  measured x-ray spectra. The theoretical uncertainty is quoted to be about
  twice as large as the experimental error.
  Besides, the errors on $b_{0}$ and on $b_{1}$ are strongly correlated.
  \par
  Using \Ref{a30} as our input, we have calculated the $\pi$d scattering
  length and the results  are presented in Table \ref{table1}.
  All entries are doubled because
  we employ two models of NN interaction: the numbers without brackets
  have been obtained using PEST wave function
  and, respectively,  the bracketed quantities
  correspond to the Yamaguchi potential.
  For each set of input values of $(b_{0}, b_{1})$ we computed $a_{\pi d}$
  using five different methods discussed before, 
  beginning from the simplest second order
  formula \Ref{a12}, through  the static model \Ref{a10} and \Ref{a26},
  up to the full Faddeev calculation without, and, with $\Delta$,
  respectively. The results of the Faddeev calculation with s-wave  
  interaction only (without $\Delta$)  
  constitute a benchmark for the various approximations.
  Contrary to what has been often claimed in the literature, the second
  order formula is insufficient as the error incurred is roughly
  four times bigger than the present experimental uncertainty on $a_{\pi d}$.
  It is apparent from Table \ref{table1}  that the closest to 
  Faddeev result is in all cases the static model \Ref{a10}.
  The accuracy of the latter is very good, the error being 
  always below 2\%. By contrast,
  the performance of the implementation \Ref{a26} of static model 
  is rather disappointing, especially that
  from formula \Ref{a26} containing binding energy correction,
  one might expect further improvement. Nevertheless, the numbers 
  show just the opposite, that in fact the included corrections
  go in the wrong  direction worsening the results so much
  that even the second order formula proves to be more accurate.
  Of course, it is not just the binding energy correction
  that is responsible
  for the difference between the static model and the Faddeev result,
  as only the latter properly accounts for the nucleon recoil. 
  However, the lions share of the recoil correction seems to be 
  cancelled with the binding energy correction and
  this cancellation explains the success of
  the static formula \Ref{a10}  
  containing neither of these corrections. An explicit demonstration
  that, at least to the second order,  
  such mechanism is at work can be found in Ref. \cite{FAL77}.
  \par 
   Since the static model \Ref{a10} proves to be so accurate for
   Yamaguchi and PEST models of the NN interaction, we took advantage
   of this fact, using it to examine more realistic NN potentials containing
   also the D-wave part. The results of our computations are displayed
   in  table \ref{table2} where we compare the two separable models
   (Hulthen-Yamaguchi and PEST), used in Faddeev
   calculations, with two popular local potentials 
   (Paris \cite{Paris} and Bonn \cite{Bonn}). As expected, 
   the PEST wave function results are indeed very
   close to those obtained with  Paris wave function despite the lack of
   the D-component in the PEST wave function. Therefore, 
   neglecting the D-wave in the Faddeev calculation does not
   appear to be a serious omission. 
   It is also gratifying that PEST, Paris and Bonn models
   give very similar results.  
   \par
   In table \ref{table3} we present the values of $\pi$d scattering
   length obtained in result of iterative solution of the Faddeev
   equations. Since for zero-range $\pi$N interaction  there is no
   additional suppression due to the $\pi$N  form factor,
   the rate of convergence is somewhat slower but the converged
   result is obtained in less than 10 iterations. 
   We give  $a_{\pi d}$
   values calculated  with and without p-wave $\pi$N interaction what
   allows to evaluate the p-wave contribution in each order.
   For Yamaguchi NN interaction the p-wave correction in the first order
   is  quite large and  contributes $0.47 \times 10^{-2}/m_{\pi}$.
   The p-wave contribution to the second order 
   (called sp-term in Ref. \cite{BARU})
   has  opposite sign and equals $-0.35\times 10^{-2}/m_{\pi}$. 
   In general, the net effect of the p-wave interaction 
   on the converged result is reduced  owing to the
   destructive interference between repulsive s-waves and attractive
   p-waves, amounting in total only $0.29 \times 10^{-2}/m_{\pi}$.
   Similar features are observed for the PEST model but 
   since the convergence rate is faster, the higher order corrections
   are suppressed and the
   interference effects seem to be smaller, i.e. the first order
   p-wave correction is $0.45\times 10^{-2}/m_{\pi}$ while the corresponding
   correction to the converged result is $0.39\times 10^{-2}/m_{\pi}$. 
  \par
  It is apparent form table \ref{table1} that
  the calculated $\pi$d scattering length values are rather sensitive
  to the input values of $(b_{0}, b_{1})$ and therefore it is not
  so easy to see when the calculation agrees with experiment. 
  To facilitate the comparison with experiment
  the values of $a_{\pi d}$
  resulting from Faddeev calculation (PEST with $\Delta$)
  and displayed in table \ref{table1} have been
  represented analytically using bilinear interpolation on a
  grid in the $(b_{0}, b_{1})$ plane. 
  Then, given the interpolating polynomial, we equated it to
  the experimental value of $a_{\pi d}$, adding 
  or subtracting the experimental error. This 
  procedure gave us two constraints
  of algebraic form in the  $(b_{0}, b_{1})$ variables, 
  readily solvable with respect to one of these variables.
  The two functions obtained this way  may be plotted
  in the $(b_{0}, b_{1})$ plane   
  where, as shown in Fig. \ref{fig1}
  they set the boundary of the  
  tilted band representing the one
   standard deviation constraint imposed by the $\pi d$ 
   scattering length deduced form pionic deuterium data.  
  The rectangle  in Fig. \ref{fig1} represents the experimental values of  
   $(b_{0}, b_{1})$ to within one standard deviation inferred from
   pionic hydrogen data.
   The ultimate $(b_{0},b_{1})$ values that are consistent with both the
   pionic hydrogen and the pionic deuterium data fill the area of the
   black strip.

   \section{Finite range approach}
   \label{se:three}
   
   Thus far our treatment of the pion-deuteron scattering problem
   has been carried out exclusively within the zero-range model.
   Although this model has served us well, it is based on certain
   idealization whose validity and consequences need to be examined. 
   We therefore turn now to the question of formulating a finite range
   version of the approach presented in the preceding section.
   Relaxing the zero-range limitation has of course its
   {\it quid pro quo} in that we have to worry now about the off-shell
   extension of the $\pi$N scattering amplitude and this means that
   the pionic hydrogen problem has to be considered
   {\it ab intitio} in order to provide the
   necessary input for the $\pi$d calculation. Anticipating the
   application in the Faddeev type calculation, it will be
   convenient for us to work with separable potentials. To get
   insight into the pionic hydrogen problem, 
   let us consider a two-channel situation, where 
   the upper channel labeled as 1 corresponds to the neutral $\pi^{0}n$ system
   and the lower channel labeled as 2, respectively, to the
   $\pi^{-}p$ system. We assume that the two-channel interaction
   respects isospin invariance and the isospin symmetry is broken only by the 
   Coulomb potential operative 
   in channel 2 and by the mass splitting within
   isospin multiplets.  Since we wish to consider an atomic system  
   it is essential to treat the Coulomb interaction exactly.
   To meet this requirement, we choose the two-channel
   Lippmann-Schwinger equation as our dynamical framework that in 
   coordinate representation takes the form
   \begin{subequations}
   \label{b0}
   \begin{equation}
   u_{1}(r)=\int_{0}^{\infty}
   \bra r \left | G_{1}^{+}(W) \right | r'\ket \,
   \left [ V_{11}(r',r'')\,u_{1}(r'') + V_{12}(r',r'')\,u_{2}(r'')\right ]
   \,\D r'\;\D r'' 
   \label{b0:a}
   \end{equation}
   \begin{equation}
   u_{2}(r)=\int_{0}^{\infty}
   \bra r \left | G_{2}^{+}(W) \right | r'\ket \,
   \left [ V_{21}(r',r'')\,u_{1}(r'') + V_{22}(r',r'')\,u_{2}(r'')\right ]
   \,\D r'\;\D r'' 
   \label{b0:b}
   \end{equation}
   \end{subequations}
   where we have assumed spherical symmetry of the problem and
   $u_{j}(r)$ denotes  zero orbital momentum
   radial wave function in channel $j$ .
   The strong $\pi N$ interaction is adopted here in the form of
   a non-local potential matrix $V_{ij}$. In \Ref{b0} we have
   introduced the Green matrix whose only non-vanishing diagonal elements are
   \begin{equation}
   \bra r \left | G_{1}^{+}(W) \right | r'\ket=-(2\mu_{1} /p_{1})
   \exp{(\imath p_{1}r_{>})}\;\sin{( p_{1}r_{<})},
   \label{b2}
   \end{equation}
   for the neutral channel, while in the charged channel we have 
   to take into account the Coulomb interaction and the 
   exact Green's function in this case reads
   \begin{equation}
   \bra r \left | G_{2}^{+}(W) \right | r'\ket=-(2\mu_{2} /p_{2})
   \left [G_{0}(\eta,p_{2}\, r_{>})+\imath
   \, F_{0}(\eta,p_{2}\, r_{>})\right ] \; F_{0}(\eta, p_{2}\,r_{<}),
   \label{b3}
   \end{equation}
   where  $r_{<}=\min(r,r'),\;r_{>}=\max(r,r')$.
   In \Ref{b0}-\Ref{b3} $W$ denotes the total  c.m. energy
   (including the rest mass),  $\mu_{j}$ are
   the reduced masses in the two channels and $p_{j}$ are the 
   channel momenta: $ p_{j}=\pm \sqrt{ 2\mu_{j} (W-E_{j}) } $ with
   $E_{j}$ being the threshold energies and the sign ambiguity will
   be resolved in a moment.  All masses here are assumed to take
   their physical values. In \Ref{b3} $\eta = -\alpha \mu_{2}/p_{2}$
   and 
   $G_{0}, F_{0}$ denote the standard Coulomb wave functions for
   orbital momentum  $\ell=0$, defined in \cite{Abramowitz}.
   Finally, it should be noted that there is no
   ingoing wave in \Ref{b0}, as appropriate for a bound state problem. 
   \par
   As mentioned above,
   to simplify matters, we assume that the interaction is separable, i.e.
   that the potential matrix is 
   \begin{equation}
   V_{ij}(r,r') = -v(r)\,s_{ij}\,v(r'),
   \label{b4}
   \end{equation}
   where the function $v(r)$ represents the shape of the potential and
   the dimensionless parameters $s_{ij}$ are the measure of
   the strength of the
   potential. Time reversal implies $s_{ij}=s_{ji}$. With separable
   potentials, the system of integral equations \Ref{b0}
   can be solved analytically. To this end it is
   sufficient to multiply each of the  equations by $v(r)$ and 
   integrate over $r$. This gives a system of two homogeneous algebraic
   equations for the two unknown quantities
   $$
   X_{j} = \int_{0}^{\infty}v(r)\,u_{j}(r)\, \D r, \quad j=1,2
   $$
   and the latter will have a non-trivial solution if, and only if,
   the determinant of the system $D(W)$ vanishes. Expanding the determinant,
   we are led to  the explicit bound state condition
   \begin{equation}
   D(W) = \left (1+s_{11}\bra v |G_{1}^{+}(W)|v \ket \right )
             \left (1+s_{22}\bra v |G_{2}^{+}(W)|v \ket \right )
    - s_{12}^{2}\; \bra v |G_{1}^{+}(W)|v \ket \; 
    \bra v |G_{2}^{+}(W)|v \ket=0 
   \label{b5}
   \end{equation}
   where we have introduced the abbreviation
   $$
    \bra v |G_{j}^{+}(W)|v \ket=\int_{0}^{\infty} 
    v(r)\,   \bra r \left | G_{j}^{+}(W) \right | r'\ket
    \, v(r') \;\D r\; \D r'.
    $$
  The determinant can vanish only at some 
   particular value of the energy $W=E_{B}$ that will be interpreted as the
   bound state energy.
    Normally, knowing the underlying interaction, by solving \Ref{b5}
    one obtains the binding energy. However, in the problem at issue we
    have a reversed situation: we know the binding energy from
    experiment and it is the interaction that we are after.  In the case of
    the pionic hydrogen atom  we have an unstable bound state in the
    charged channel and the binding energy will be a complex number.
    We set
    \begin{equation}
    \label{b50}
    E_{B} = E_{2} + E_{1s} - (\epsilon + \imath \Half \gamma)
    \end{equation}
    where $E_{1s}=-\mu_{2}\alpha^{2}/2$ 
    is the purely Coulombic 1s state binding energy. Since in our
    formalism there is no room for the radiative decay of the pionic hydrogen  
    the partial width $\gamma$ is a fraction of
    the total width $\Gamma$ given by the formula $\gamma=\Gamma/(1+P^{-1})$
    where $P$ is the Panofsky ratio.  It has been shown in ref.
    \cite{GIB86}
    that the effect of the ($\pi^{-}$,$\gamma$) reaction on the 
    accounted for hadronic channels is negligible.
    The experimental values for
    $\epsilon, \Gamma$ (cf. \cite{Sigg}) 
    and P (cf. \cite{Panofsky}) adopted in this work, are
    \begin{eqnarray*}
    \epsilon &=&7.108 \pm 0.013\text{(stat)}\pm 0.034\text{(syst)}\,eV,\\
    \Gamma &=& 0.868 \pm 0.040\text{(stat)} \pm 0.038\text{(syst)}\,eV,\\
    P &=& 1.546\pm 0.009,
    \end{eqnarray*}
    and in the following we shall take
    $\epsilon=7.108\pm 0.047\,eV$ and $\gamma=0.527\pm 0.047\,eV$
    as the input values. 
    It must be immediately explained here  that in this work 
    we have defined  $\epsilon$ in accordance with a different convention, 
    so that our $\epsilon$ has opposite sign than that used in ref. 
    \cite{Sigg}.
    In our approach we have tacitly assumed that
    under perturbative treatment
    all electromagnetic corrections contribute the same amount to the
    purely Coulombic level and to the level shifted by strong
    interaction. More precisely, we are going to ignore the small
    effects caused by the distortion of the wave function. Accordingly,
    the electromagnetic corrections need not concern us here and they
    have been left out altogether but, of course, they would be 
    indispensable for calculating the total displacement of the level from
    its Coulombic position.
    \par
    The pole of the T-matrix that corresponds to the solution of \Ref{b5}
    can be located on one of the four Riemann sheets as appropriate for
    a two-channel problem.   This is also apparent from the
    mentioned above sign ambiguity
    in the definition of the channel momenta in \Ref{b3}.
    The right choice of the Riemann sheet is
    essential and this can be accomplished by  proper adjustment of
    the signs of the imaginary parts of the channel momenta $p_{j}$.
    We are using here the standard enumeration of the Riemann sheets
    introduced in ref. \cite{HENDRY}, i.e.
    \begin{eqnarray*}
    \text{sheet I:}   \quad & Im\, p_{1}>0;& \;  Im\, p_{2}>0\\ 
    \text{sheet II:}  \quad & Im\, p_{1}<0;& \;  Im\, p_{2}>0\\
    \text{sheet III:} \quad & Im\, p_{1}<0;& \;  Im\, p_{2} <0\\
    \text{sheet IV:}  \quad & Im\, p_{1}>0;& \;  Im\, p_{2}<0.
    \end{eqnarray*}
    In the pionic hydrogen case, with an unstable bound state in channel 2,
    we have to enforce the pole to be located on the second sheet.
    \par
  To proceed further we need some concrete shape factor $v(r)$ and our 
  choice here is the exponential shape, i.e. we set
  \begin{equation}
  v(r)=\sqrt{\beta^{3}/\mu}\; \exp{(-\beta\,r)}
  \label{b6}
  \end{equation}
  where $\mu$ is the reduced pion-nucleon mass in the case of exact isospin
  symmetry (we take average mass for each isospin multiplet) and 
 $\beta$ is the inverse range parameter.
 With the exponential  form \Ref{b6}, the potential \Ref{b4} is 
 identical with the familiar Yamaguchi potential 
 and the Green's function matrix elements can be obtained in
 an analytic form. The final result is
 \begin{equation}
    \bra v |G_{1}^{+}(W)|v \ket=-\frac{\mu_{1}}{\mu}
   \; \frac{1}{(1-\imath p_{1}/\beta)^{2}}
 \label{b7}
 \end{equation}
 for the neutral channel, while the corresponding formula for the
 charged channel reads
 \begin{equation}
    \bra v |G_{2}^{+}(W)|v \ket=-\frac{\mu_{2}}{\mu}
   \; \frac{1}{(1-\imath p_{2}/\beta)^{2}}
   \; \frac{\mbox{}_{2}F_{1}(1,\,\imath \eta;\,\imath \eta+2;\,z^{2})}
    {\imath \eta+1}
 \label{b8}
 \end{equation}
 with $z=(\beta+\imath p_{2})/(\beta-\imath p_{2})$.  
 The last fraction in \Ref{b8} accounts for the Coulomb interaction and
 the symbol
 $\mbox{}_{2}F_{1}(a,b;c;z)$  denotes the hypergeometric function
 defined in \cite{Abramowitz}.
 The computation of the hypergeometric function entering 
 \Ref{b8} is greatly simplified 
 owing to the fact that the first parameter is equal to unity 
 in which case the 
 continued fraction representation of $\mbox{}_{2}F_{1}(1,b;c;z)$
 discovered by Gauss \cite{Gauss} proves to be useful.
 The continued fraction summation converges
 in the whole of the complex  $z$ plane away form the branch cut
 on the real axis running from one to infinity.
 \par
 With exact isospin symmetry the three strength parameters $s_{11},
 s_{12}, s_{22}$ are not independent and can be expressed in terms of
 isospin 1/2 and isospin 3/2 strengths denoted hereafter
 as $s_{1}$ and $s_{3}$, respectively.
 In the bound state condition \Ref{b5} both the real and the imaginary
 part of $D(E_{B})$ have to vanish simultaneously and that gives us two
 real equations. Since the bound state energy is known
 (cf. \Ref{b50}), we put
 $ s_{11}= (s_{1}+2s_{3})/3\; ; s_{22}= (2s_{1}+s_{3})/3\; ;
 s_{12}=  \sqrt{2}(s_{3}-s_{1})/3 $
 in \Ref{b5}, and regarding $s_{1}$ and $s_{3}$
 as our two unknowns,
 we arrive at two algebraic  equations that can be solved analytically   
 \begin{subequations}
 \label{b55}
 \begin{equation}
 s_{1}^{2}\,\text{Im}\,ac^{*}+s_{1}\,\text{Im}(ab^{*}-c)-\text{Im}\,b=0;
 \label{b55:a}
 \end{equation}
 \begin{equation}
 s_{3}=-(1+s_{1}\,\text{Re}\,a)/(\text{Re}\,b+s_{1}\,\text{Re}\,c),
 \label{b55:b}
 \end{equation}
 \end{subequations}
 where $a=(\bra v|G_{1}^{+}(E_{B})|v\ket 
 +2\bra v|G_{2}^{+}(E_{B})|v\ket)/3;
 b=(2\bra v|G_{1}^{+}(E_{B})|v\ket +\bra v|G_{2}^{+}(E_{B})|v\ket)/3$ and
  $c=\bra v|G_{1}^{+}(E_{B})|v\ket \bra v|G_{2}^{+}(E_{B})|v\ket$.
 With $s_{1}$ and $s_{3}$ in hand, the  corresponding
 scattering lengths ($a_{2I}$ with I=1/2 and 3/2) are obtained from
 \begin{equation}
 a_{2I}=\frac{2}{\beta} \frac{s_{2I}}{1-s_{2I}}.
 \label{b10}
 \end{equation}
 \par
 For local potentials the method outlined above could be
 also applied but in such case it would be more
 convenient to use instead of \Ref{b0} an equivalent
 set of two coupled Schr\"odinger equations. For fixed energy
 and the proper choice of the Riemann sheet,
 these differential equations can be integrated numerically and the 
 bound state equation is obtained from the requirement of vanishing
 of the Wronskian determinant. The latter is again a function of the 
 isospin 1/2 and isospin 3/2 strength parameters, or
 if one prefers, the corresponding potential
 depths. Although the bound state condition is   
 defined then only numerically but from it one can get
 two real equations that can be solved numerically using standard procedures.
 With a local potential, however, the solution of the three-body
 problem becomes much more complicated and this is the main reason
 why we preferred to work with a separable potential.
 \par
 Our calculational scheme is now complete 
 and we shall present our results.  Using as our input 
 the experimental values of the pionic hydrogen level shift and width,
 the bound state equation has been solved analytically 
 by adopting a number of ''reasonable'' values for $\beta$ and 
 in our computations we have used the
 values from $2\,fm^{-1}$ to $10\,fm^{-1}$. Although,
 we do not know the precise value of the range but 
 there is no physical mechanism known that might generate 
 long range forces in the $\pi N$ system,
 the longest range is unlikely to be bigger than $0.5\,fm$ and
 this sets the lower limit of acceptable $\beta$ values. In principle,
 there is no upper limit for $\beta$ but for $\beta>10\,fm^{-1}$ we have
 in practice reached the limit of the zero-range forces
 and things change very little above that limit.
 The exact solutions of the bound state equation are presented in 
 Table \ref{table4} 
 where the errors reflect only the experimental uncertainty of
 our input, i.e. $\epsilon, \Gamma$ and the Panofsky  ratio.
 Our isoscalar and isovector scattering lengths are in
 good agreement with the values extracted in \cite{Sigg}. 
 This has been illustrated in Fig. \ref{fig3} where we have compared
 a representative sample of 
 our computations with the values obtained by 
 Sigg {\it et al.} \cite{Sigg}.
 The solutions corresponding to $\beta$ spanning the range 
 $2-10$ fm$^{-1}$
 are located very close to each other in the (b$_{1}$, b$_{0}$) plane
 and putting more than three points on the plot might have obscured
 the picture. The error bars reflect only the experimental
 uncertainty of our input.
 As mentioned above, the bound state equation \Ref{b5} 
 yields a second order equation for $s_{1}$ and $s_{3}$ and therefore
 we have always two solutions (cf. \Ref{b55}). 
 Only one of them is presented in Table \ref{table4} whereas
 the second solution leads to both $b_{0}$ and $b_{1}$ positive and
 has had to be rejected. 
 When the two strength parameters $s_{1}$ and $s_{3}$ are known
 we can calculate not only the scattering lengths 
 but also the effective ranges
 in each of the two isospin states and these values are presented 
 in Table \ref{table4}.  Instead of the effective range we
 use the parameter $B_{2I}$ that is defined from the
 expansion of the real part of the s-wave scattering amplitude
 in powers of the c.m. momentum $k$, i.e. close to threshold, we have 
 $Re f_{2I}(k)=a_{2I}+B_{2I}\,k^{2}+\cdots$.
 For comparison,  at the bottom of Table
 \ref{table4} we give the values of all parameters inferred from
 a recent phase shift analysis \cite{Gashi}.
 The calculated scattering lengths, listed in Table \ref{table4}, 
 are almost independent upon $\beta$, in contrast with the slope
 parameters $B_{2I}$ which change quite a bit when $\beta$
 is varied in the interval 2-10 fm$^{-1}$. In addition to that,
 our  $B_{3}$ values
 turn out to be always positive and therefore have opposite sign 
 than those deduced from phase shift analysis \cite{GIB98,Gashi}.
 Actually, the pionic hydrogen data  
 provide a strong constraint only for the scattering lengths and
 sticking to a simple $\pi$N Yamaguchi potential
 it is not possible to get $B_{3}$ negative just 
 by varying $\beta$. Indeed, for fixed a$_{2I}$ 
 the slope parameter B$_{2I}$ 
 is given by the exact formula
 $$
 B_{2I}=-a_{2I}^{3}\;[1+\frac{1}{2\beta a_{2I}}(3+\frac{4}{\beta
 a_{2I}})]
 $$
 and since the expression in the square bracket is necessarily positive
 the sign of B$_{2I}$ is bound to be opposite to that of a$_{2I}$.
 To obtain a negative B$_{3}$ a more sophisticated
 potential involving both repulsion and attraction would have been required
 \cite{GIB98}. 
 There is no need for such extension, however, because our model
 has been devised  for describing only the near threshold phenomena
 and is quite adequate at that.  Expanding the phase shift
 close to threshold in powers of $k$, we have 
 $\delta_{2I}=a_{2I}k+O(k^{3})$ and it is apparent that a model
 providing merely the scattering length reproduces satisfactorily
 the phase shift in the neighbourhood of zero where $\delta_{2I}$
 exhibits a linear behaviour.  
 In our case this is all that counts  as we never
 deal with higher energies. This means that  
 the determination of the slope parameters is out of reach
 within our model since the appropriate
 energy scale has been set by the Coulomb energy in the
 pionic hydrogen, in which case terms proportional to $B_{2I}$
 make negligible contribution. For an assigned value of $\beta$
 the slope parameters may  be calculated but they are of no
 physical significance and comparing them with those resulting
 from phase shift analysis does not make much sense.
 \par
 As noted in \cite{Sigg,GIB86},
 at the energy value close to the unstable bound state
 in channel 2, the scattering
 amplitude in the open channel 1, shows a strong resonant behaviour.
 For a separable potential, the s-wave scattering amplitude $f(W)$
 in channel 1  may be easily
 calculated analytically and takes a simple form 
 \begin{equation}
 f(W)= \text{e}^{ \I \delta } \sin \, \delta /p_{1}=
 \frac{\mu_{1} } {\mu} \frac{2}{\beta} \quad
 \frac{  s_{11}+ (s_{11}s_{22}-s_{12}^{2})
 \bra v |G_{2}^{+}(W)|v \ket } {(1+p_{1}^{2}/\beta^{2} )^{2}\;D(W) },
 \label{b16}
 \end{equation}
 where $\delta$ is the corresponding phase shift that for real $W$
 below the $\pi^{-}$p threshold  is a real number.  
 The resonance is not of a Breit-Wigner shape but its position $E_{r}$ 
 may be easily established from \Ref{b16} as the energy at which
 the phase shift is equal to $\Half\pi$. Close to the
 resonant energy, i.e. for $W\approx E_{r}$ we have 
 $\cot \delta \approx (W-E_{r})/(\Half\Gamma_{r})$ and this allows
 us to infer the value of the width $\Gamma_{r}$ of the resonance.
 In ref. \cite{Sigg} the values of $(\epsilon,\gamma)$ have 
 been calculated by identifying them with $(E_{2}+E_{1s}-E_{r},\Gamma_{r})$.
 In principle, the values of $(\epsilon,\gamma)$  
 obtained that way do not have to be identical with those determined
 from the position of the bound state pole. To check that point,
 we have repeated the procedure of ref. \cite{Sigg}
 but using our separable potentials
 whose depths have been adjusted to reproduce the values
 of $(\epsilon,\gamma)$  obtained in  \cite{Sigg}. 
 We found that the two methods give nearly identical results and
 the differences in $(\epsilon,\gamma)$ did not exceed 1 meV.
 For illustration, in Fig.\ref{fig3} we show the behaviour of $\sin \delta$
 close to the resonance for the case of $\beta=3\,fm^{-1}$ where
 the strengths parameters inferred from the pole location  
 were $s_{1} =0.271820$ and $s_{3}=-0.245868$. 
 \par
 Before concluding  
 our discussion of the pionic hydrogen we wish to mention
 one last thing, namely we are going to show
 how from \Ref{b5} one can retrieve the Deser-Trueman formula
 (cf. \cite{Deser}). This task will be accomplished by obtaining an
 approximate solution of \Ref{b5} and to this end
 \Ref{b5} is cast to the form
 \begin{equation}
 1+s_{\text{eff}}(W)\; \bra v |G_{2}^{+}(W)|v \ket =0,
 \label{b11}
 \end{equation}
 where we have introduced an effective energy dependent complex strength 
 parameter $s_{\text{eff}}$,  
 defined as
 \begin{equation}
 s_{\text{eff}}(W)=s_{22}-s_{12}^{2} \bra v|G_{1}^{+}(W)|v \ket
            /\left (1+s_{11} \bra v|G_{1}^{+}(W)|v \ket \right ).
 \label{b12}
 \end{equation}
 The complex $\pi^{-} p$ scattering length $a_{\pi p}$ can be expressed
 in terms of $s_{\text{eff}}(W)$ evaluated at threshold
 \begin{equation}
 a_{\pi p}= \frac{\mu_{2}\,2}{\mu\,\beta}
 \frac{s_{\text{eff}}(E_{2})} {1-s_{\text{eff}}(E_{2})},
 \label{b13}
 \end{equation}
 and the Coulomb corrected $\pi^{-}$p scattering length, denoted as
 $a_{\pi p}^{c}$, can be obtained from the exact formula derived in
 \cite{vanH}
 \begin{equation}
 1/a_{\pi p}^{c}=\text{e}^{\xi}/a_{\pi p}+
 2\mu_{2}\alpha \; \text{Ei}(\xi),
 \label{b13a}
 \end{equation}
 where $\xi = 4 \alpha\mu_{2}/\beta$ and
 Ei($\xi$) is the exponential integral function defined in
 \cite{Abramowitz}. It should be noted here that the zero-range limit
 ($\beta\to \infty$) does not exist in \Ref{b13a} because  
 the function Ei($\xi$) for $\xi$=0 
 has a logarithmic singularity. For the case of
 $\beta=3\,fm^{-1}$ just considered, we obtain
 \begin{eqnarray*}
 a_{\pi p}    &=& 0.12081 + \imath\, 0.004441\,fm; \\
 a_{\pi p}^{c}&=& 0.12068 + \imath\, 0.004458\,fm;
 \end{eqnarray*}
 so that the Coulomb corrections do not exceed a fraction of a
 percent. However, in general,
 the Coulomb correction is model dependent, and, in particular, it is rather
 sensitive to the range of the nuclear potential what can be seen 
 when the above result is juxtaposed with the $\pi$d case  where  
 the range of the potential was comparable with the size of the deuteron 
 and, accordingly, the Coulomb correction to $\pi$d 
 scattering length was much bigger (1.5\%).
 \par
 Since  we wish to obtain an approximate solution of \Ref{b11} that
 is located not far from the Coulomb bound state, we set
 $W=E_{2}+E_{1s}+\delta E$ where $\delta E$ is a small displacement.
 To calculate $\delta E$ and
 derive the Deser-Trueman formula   
 from \Ref{b11},  we have to assume that ({\it i}) the
 complex energy shift $\delta E = -\epsilon -\imath \Half \gamma $ is small
 in comparison with Coulomb energy ($|\delta E/E_{1s}|<<1 $), and,
 ({\it ii}) that the range of the strong interaction is small as compared 
 with the Bohr radius ($\beta>>\mu_{2}\alpha $).
 Introducing a complex momentum
 $p_{c}=\sqrt{2\mu_{2}E_{1s}}=\imath \mu_{2}\alpha $
 corresponding to the Coulomb bound state, we can see that
 when $p_{2} \to p_{c}$ then $\imath \eta \to -1 $ and
 the  Green's function \Ref{b8} 
 occurring in \Ref{b11} becomes singular. 
 This singularity is of paramount importance since it induces a zero in
 the nuclear S-matrix  that is necessary to cancel the bound pole in the
 Coulomb S-matrix. As a result of this cancellation, the full S-matrix
 in the charged channel, which is a product of the Coulomb S-matrix
 and the nuclear S-matrix, remains finite at $p_{2}=p_{c}$.
 In compliance with the small shift assumption, we set  
 $p_{2}=p_{c}+\delta p $ where $\delta p $ is supposed to be a small 
 correction and since the most rapid variation  in \Ref{b8} 
 arises on account of the pole term,  we approximate
 $1+\imath \eta$ by $\delta p/p_{c} $. Apart from
 that, elsewhere we replace $p_{2}$ by $p_{c}$.
 The hypergeometric function
 for $\imath\eta = -1$ reduces to a polynomial $1-z^{2}$
 and neglecting small terms
 of the order of $p_{c}/\beta $, from \Ref{b11} we obtain  
 \begin{equation}
 \delta p \approx -4\imath\,(p_{c}^{2}/\beta)\,(\mu_{2}/\mu) 
 s_{\text{eff}}(E_{2}) \approx -2\imath\, p_{c}^{2}\,a_{\pi p} 
 \label{b14}
 \end{equation}
 where we have used \Ref{b13}
 retaining only linear term in $a_{\pi p} $. The above result
 gives the Deser-Trueman formula \cite{Deser} in its standard form
 \begin{equation}
 \delta E \approx  p_{c}\, \delta p/\mu_{2} \approx
 -2\mu_{2}^{2}\,\alpha^{3}\, a_{\pi p},
 \label{b15}
 \end{equation}
 where, in view of the above discussion, it does not really matter whether
 we take $a_{\pi p}$ or $a^{c}_{\pi p}$.
 It is perhaps in order to recall that although the Deser-Trueman
 formula \Ref{b15} has been derived here for a specific choice of
 the underlying interaction, but its validity is quite general.
 To examine the accuracy of Deser-Trueman formula we turn again to our
 previous example when $\beta=3\,fm^{-1}$ and by computing $a_{\pi p}$
 from \Ref{b13} and inserting in
 \Ref{b15}, we obtain
 $(\epsilon, \gamma)$ = (7.024, 0.516) eV to be compared with our input
 values equal $(\epsilon, \gamma)$ = (7.108, 0.527) eV that ought
 to have been reproduced if formula \Ref{b15} had been exact.
 It is a remarkable property of the Deser-Trueman formula  that it is
 independent of the range of the underlying interaction and therefore
 the error in this formula must be of the same size as the uncertainty
 in the exact result caused by varying $\beta$. If one is prepared
 to tolerate such uncertainty formula \Ref{b15} could be used to
 infer $a_{1}$ and $a_{3}$. Introducing a two-channel K-matrix,
 isospin invariance can be invoked to pin down its
 elements at the single unsplit threshold 
 $$
 K=\left (\;
 \begin{matrix}
 & \tfrac{1}{3}a_{1}+\tfrac{2}{3}a_{3}
 & \tfrac{\sqrt{2}}{3}(a_{3}-a_{1})\\
 & \tfrac{\sqrt{2}}{3}(a_{3}-a_{1})
 & \tfrac{2}{3} a_{1}+\tfrac{1}{3} a_{3}\\
 \end{matrix}
 \;\right )
 $$
 and the complex $\pi^{-}p$ scattering length takes the form
 \begin{equation}
 a_{\pi p}= K_{22}+\imath p_{t}K_{12}^{2}/
 (1-\imath p_{t}K_{11}),
 \label{b30}
 \end{equation}
 where $p_{t}$ is the momentum in the $\pi^{0}n$ channel evaluated at
 the $\pi^{-}p$ threshold.  The scattering length \Ref{b30}, 
 unlike \Ref{b13}, does not depend upon the range. 
 Inserting \Ref{b30} in \Ref{b15}
 and separating the real and the imaginary part, we end up with
 two real equations for the two unknowns $a_{1}$ and $a_{3}$.
 To more than sufficient accuracy, the explicit solutions, are
 \begin{subequations}
 \label{b31}
 \begin{eqnarray}
 a_{1}&=&\left [x \pm y(1-2p_{t}y)/\sqrt{2p_{t}y}\right ]/(1-p_{t}y);
 \label{b31:a}\\
 a_{3}&=&\left [ x \mp y(2-p_{t}y)/\sqrt{2p_{t}y}\right ]/(1-p_{t}y),
 \label{b31:b}
 \end{eqnarray}
 \end{subequations}
 where $x=\epsilon/2\mu_{2}^{2}\alpha^{3},
 y=\Half\gamma/2\mu_{2}^{2}\alpha^{3}$ and the double sign in \Ref{b31} 
 stems the fact that eq. \Ref{b30} is quadratic in $a_{2I}$.
 If ($\epsilon$,$\gamma$)
 have been obtained in a model independent way then the results \Ref{b31} 
 are also model independent.
 As seen from Table \ref{table4} the uncertainty on $a_{1}$ and
 $a_{3}$ (3\% and 9\%, respectively) induced by experimental errors 
 on $(\epsilon, \gamma)$ is much bigger than the uncertainty caused by 
 varying $\beta$ (about 1\%).
 Under these circumstances it is perfectly justified 
 to infer the $\pi$N scattering lengths
 via Deser-Trueman formula and their numerical values
 obtained from \Ref{b31} are displayed in Table \ref{table4} whereas 
 the corresponding $b_{0}$ and $b_{1}$ are presented in Fig. \ref{fig3}.
 \par
 It is apparent from Table \ref{table4} that to improve upon
 Deser-Trueman formula we need
 some additional clue concerning $\beta$ and it becomes something
 of a challenge to find ways to ferret out more precisely what the
 value of $\beta$ might be. 
 So far in our considerations we have not mentioned yet the pionic deuterium
 data and at this stage it is logical to ask whether this additional
 information might not help to pin down the 
 range parameter of the $\pi$N potential. Therefore,
 in the next step, we use the values given in Table \ref{table4}
 as input for a
 three-body calculation, i.e. we use the separable potential \Ref{b6}
 in the Faddeev equations
 for calculating the $\pi d$ scattering length. The results of our
 computations are presented in Fig. \ref{fig4} where we have plotted  the 
 $\pi d$ scattering length versus $\beta$. The full circles represent
 the results obtained by the including the p-wave interaction
 (more precisely, just the P33 wave), while
 the open circles correspond to a situation where the delta has been 
 left out. For reasons of clarity of the presentation these two sets
 of points have been given at
 different $\beta$ values.
 The indicated error bars reflect the uncertainty in the
 input values (cf. Table \ref{table4} ). 
 For comparison, the experimental value of
 $\pi d$ scattering length to within one standard deviation 
 is given in Fig \ref{fig4}  
 as the area between the two horizontal lines.  The striking
 feature apparent from Fig \ref{fig4} 
 is that the results are almost independent
 of the range parameter $\beta$.  Furthermore, the calculated 
 scattering lengths are consistent with experiment for all $\beta$
 no matter whether the delta has been included or not. 
 This result may come as a disappointment since the deuteron data
 give no illumination how to bracket the value of  $\beta$.   
 \par
 In order to understand how the above result comes about we shall
 invoke again the static model, taking advantage of the fact that
 with the Yamaguchi potential representing the $\pi$N 
 interaction the static solution
 of the Faddeev equations may be readily obtained (cf. ref.\cite{SE} ).
 Thus, introducing
 the Yamaguchi form factors and going to the static limit we can repeat
 the procedure outlined in the preceding section. The static solution
 of the Faddeev equations may be then sought in the form
 \begin{eqnarray*}
 A(\bbox{q},\bbox{k})&=&
 -\frac{m}{2\pi}\;\frac{\beta^{2}}{q^{2}+\beta^{2}}\;{\cal A}(k);\\
 X(\bbox{q},\bbox{k})&=&
 -\frac{m}{2\pi}\;\frac{\beta^{2}}{q^{2}+\beta^{2}}\;{\cal X}(k),
 \end{eqnarray*}
 and the above ansatz used in the Faddeev equations yields a set of
 two integral equations that differ from \Ref{a23} in that the
 appropriate kernels contain now an extra factor 
 $1/[1+(\bbox{k}+\bbox{k'})^{2}/\beta^{2}]^{2}$. Despite this
 additional complication, the
 Fourier transform of this extended kernel still can be effected 
 and leads to a simple analytic expression
 $$
 \frac{4\pi}{\kappa^{2}+(\bbox{k}+\bbox{k'})^{2}}
 \frac{\beta^{4}}{[\beta^{2}+(\bbox{k}+\bbox{k}')^{2}]^{2}}=
 (1-\frac{\kappa^{2}}{\beta^{2}})^{-2} \int
 e^{-\imath(\bbox{k}+\bbox{k}')\bbox{r}} \frac{\D^{3}\,r}{r}
 \left \{e^{-\kappa r}-e^{-\beta r}[1+\frac{\beta r}{2}
 (1-\frac{\kappa^{2}}{\beta^{2}})] \right \}.
 $$
 Using the above formula, 
 similarly as before, we end up with a system of two algebraic
 equations for $A(r)$ and $X(r)$. Neglecting the binding energy correction
 ($\kappa \to 0$), the resulting equations differ from \Ref{a8} in that
 the zero-range pion propagator $1/r$ has to be multiplied by 
 the function $g(r)$ given by the formula
 \begin{equation}
 \label{b17} 
 g(r)= 1- e^{-\beta r}(1+\Half \beta r).
 \end{equation}
 Therefore,  the sought for solution for $A(r)$ follows from \Ref{a10}
 after replacing $1/r$ by $g(r)/r$. Formula \Ref{b17} proves to be quite
 useful for estimating the size of the $\beta$ dependent
 correction and to this end we need to evaluate $g(r)$ at some
 average value of $r$ and a plausible candidate for such average value
 is the deuteron radius 
 $r_{d}=\Half\sqrt{<r^{2}>}\approx 2$ fm. 
 Indeed, with this choice the second order formula \Ref{a12} that provides
 for a major contribution to $a_{\pi d}$ will be little affected since 
 by setting $r=r_{d}$,
 we get $\bra 1/r \ket$ = 0.5 fm$^{-1}$, not far from the values listed
 in Table \ref{table2}. When $\beta$ is varied in
 the range $2-10$ fm$^{-1}$, we have r$_{d}\beta >$4
 in the exponential damping factor in \Ref{b17}, so that the 
 $\beta$ dependent terms make a contribution to $g(r)$ at the level of
 a few percent and the resulting $\pi$d scattering length is almost
 independent upon $\beta$. This feature, sustained in the full
 Faddeev solution, is a consequence of the fact that the adopted range
 of the $\pi$N forces was small as compared with the deuteron radius.
 \par
 In conclusion, we have seen that
 the uncertainty in the calculated $a_{1}$ and $a_{3}$,
 as well as in $a_{\pi d}$, connected with the lack of knowledge of the
 range parameter constitutes only a small fraction of the uncertainty
 resulting from the experimental errors on the pionic hydrogen data.
 The above results may be viewed as an {\it a posteriori} justification
 of our zero-range model developed in Sec. \ref{se:two}: introducing 
 a finite range would be merely a fine tuning
 which is not yet affordable in the current state of affairs. 
 
   \section{Discussion}           
   \label{se:four}                
 
 Assuming that the underlying $\pi$N interaction is isospin invariant,
 we have analysed the recent pionic hydrogen and pionic deuterium data
 with the purpose to extract from them  $\pi$N  s-wave
 scattering lengths $a_{2I}$ for I=1/2 and I=3/2.
 It is an empirical fact that the complex energy shift 
 in either of these two atomic systems is small
 when compared with the corresponding Coulomb energy and 
 with the appropriate Bohr radii setting the length scale, 
 the  $\pi$-p and $\pi$-d interactions are of a short range.    
 Under these circumstances  Deser-Trueman formula
 provides an extremely good approximation,
 relating in a model independent way  the 1s level shifts and widths 
 in the pionic hydrogen and pionic deuterium to  the complex scattering
 lengths $a_{\pi p}$ and $a_{\pi d}$, respectively. However,
 to infer $a_{2I}$ from the latter quantities is a non-trivial 
 dynamical problem and to be able to solve it we    
 introduced a simple and transparent potential representation of
 the $\pi$N interaction. 
 Within this model we obtain explicit solution      
 of the $\pi^{-}p$ bound state problem and also of the related
 three-body $\pi$d scattering problem at zero energy. 
 \par
\begin{figure}[ht]
 \centering
 \caption{
  Constrains on the isoscalar and isovector scattering lengths imposed by  
  pionic deuterium data. The black strip corresponds the one standard
  deviation region. The rectangle corresponds to the values  
  obtained from pionic hydrogen data  in \protect\cite{Sigg}.
      }
 	 \label{fig1}
  	    \vspace*{8mm}

  \includegraphics[width=0.5\textwidth,totalheight=0.5\textheight]{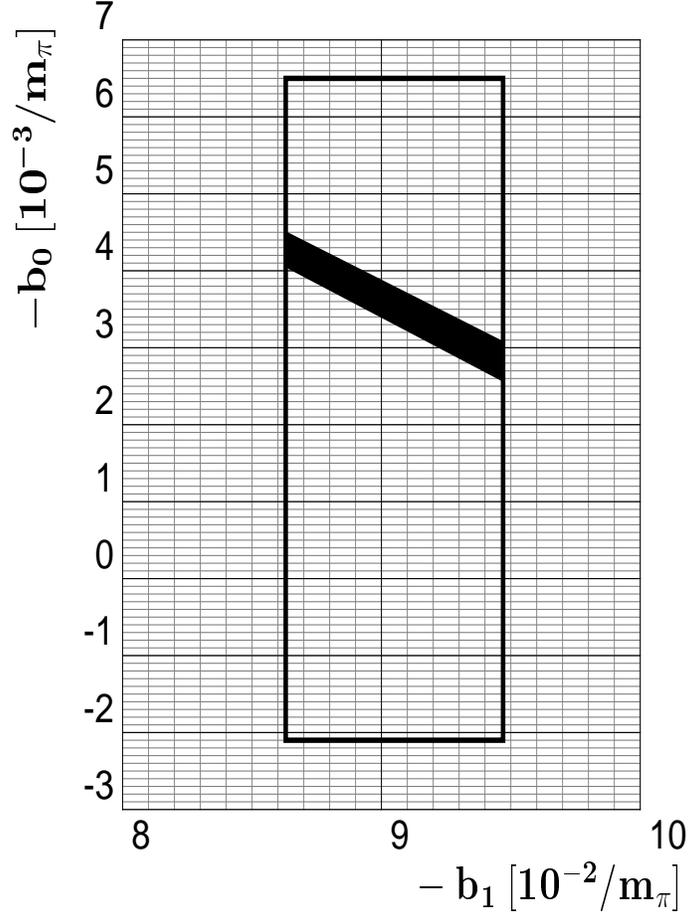}
 	       \end{figure}

 We have assumed throughout  this work
 that the $\pi$N forces are of a very
 short range and this supposition follows from a particle
 exchange picture: there is no sufficiently light particle presently known
 that might be capable of generating forces whose range would 
 exceed 0.3-0.4 fm (which roughly corresponds to a vector meson exchange).
 In this situation it was logical to take the zero-range limit
 as our point of departure.
 In order to find out what the deuteron data
 can teach us about $\pi$N scattering  lengths, 
 we calculated the $\pi$d scattering length 
 by solving the appropriate three-body $\pi$NN problem.
 This task was accomplished, both within the static
 approximation, and also by using the Faddeev formalism.
 We demonstrated that the same static formula for $a_{\pi d}$ can
 be derived from: {\it (i)} a set of boundary conditions; {\it (ii)} 
 a static solution of Faddeev equations, and {\it (iii)} 
 a summation of Feynman diagrams. The static formula 
 expressing $a_{\pi d}$ in terms of $\pi$N scattering
 lengths was found to be surprisingly accurate: 
 the error, estimated by  comparing the static result with 
 the full Faddeev solution, was at the level of 2\%,
 i.e.  of the same size as the experimental error on $a_{\pi d}$.
 The standard second order formula was shown  
 to be insufficient: the incurred error was
 three times bigger than the present experimental 
 uncertainty on $a_{\pi d}$.
 Using as input the $\pi$N scattering lengths, that had been 
 inferred earlier \cite{Sigg} from pionic hydrogen data,  
 we obtained $a_{\pi d}$ by solving    
 the  Faddeev equations for zero-range $\pi$N forces. 
 The requirement that the calculated 
 $a_{\pi d}$ be in agreement with experiment
 to within one standard deviation, imposes bounds 
 on the isoscalar and isovector $\pi$N scattering lengths.
 The values of the $\pi$N scattering lengths
 determined that way, consistent
 with both the pionic hydrogen and the pionic deuterium data,
 are presented in Fig. \ref{fig1}.
 \par
 In the next stage of this investigation we
 lifted the zero-range limitation introducing a range parameter. 
 The pionic hydrogen bound state problem was solved afresh for 
 a variety of range values.
 We derived the appropriate bound state condition and
 taking the 1s level shift and width of the pionic
 hydrogen as input, we used this condition to
 determine the s-wave $\pi$N potentials. This was possible 
 since a complex condition is equivalent to two real equations, which
 for an assigned range, can be exactly solved 
 for the I=1/2 and I=3/2 depth
 parameters entering the $\pi$N potentials.
 Knowing the potentials, it was a trivial matter to calculate 
 the corresponding s-wave scattering amplitudes.
 As can be seen from Table \ref{table4},
 the resulting $\pi$N scattering lengths are rather insensitive to the 
 adopted value of the range parameter.
 \par
 The analysis of the pionic hydrogen presented in this work parallels that
 given in \cite{Sigg}. We differ, however, in the adopted dynamical
 frameworks: in \cite{Sigg} Klein-Gordon equation together with a local
 $\pi$N potential has been used, whereas we consider a non-relativistic
 Lippmann-Schwinger equation (equivalent to a Schr\"odinger equation)
 with a separable $\pi$N potential. 
 As may be seen from Fig. \ref{fig3}, the $\pi$N scattering lengths
 inferred in this paper are in good agreement with those deduced
 in  \cite{Sigg}.  This is a direct consequence of
 the fact that Deser-Trueman formula provides such a 
 good approximation that we can make considerable progress in deducing
 the $\pi$N scattering lengths without committing ourselves in great
 deal to the nature of the $\pi$N dynamics.
 Since Deser-Trueman formula depends neither upon the shape of the $\pi$N
 potential nor upon its range, the small changes in the
 $\pi$N scattering lengths
 caused by varying the range parameter, must be attributed to the
 differences between the approximate Deser-Trueman formula and the 
 exact range dependent solutions
 of the bound state equation. Thus, Fig. \ref{fig3} illustrates the
 accuracy of Deser-Trueman formula.
 \par
 For an assigned range value, the pionic hydrogen data specify completely the 
 $\pi$N potentials,  so that they  may be used in
 the Faddeev equations in order to obtain the $\pi$d scattering length.
 The latter quantity was shown to be almost independent upon the range
 parameter (cf. Fig. \ref{fig4}) but was rather sensitive to the values of the
 $\pi$N scattering lengths used as input in the Faddeev equations. 
 The above finding, supporting the zero-range approach,  
 could be explained by the fact that 
 the range of the $\pi$N interaction that was considered physically justified 
 was small in comparison with the deuteron size.     
 \par
 We conclude that the
 lack of knowledge of the range of the $\pi$N interaction is
 responsible for some 
 uncertainty in the deduced $\pi$N scattering lengths  
 but this uncertainty is rather small, at the level of 1\%. 
 The main source of error is still the experimental
 uncertainty in the pionic hydrogen data.
 \par
 It is rather obvious that the presented model
 contains several omissions but
 we think that they are not too severe, especially that the  investigation
 has been confined to near threshold phenomena. 
 As in all non-relativistic models based on static potentials virtual particle
 production, crossing symmetry, retardation and relativistic effects
 have not been even touched upon.
 Besides that, a separable potential is not considered
 to have a strong theoretical basis and has been adopted here merely
 for convenience as it simplifies considerably the solution of
 the Faddeev equations. There are also limitations on the completeness
 of the Faddeev approach where by restriction to three-body channels
 we were forced to leave out a wealth of inelastic features. The
 absorption channels leading to two-nucleon states are not easily
 incorporated in a Faddeev theory and require considerable
 enlargement of the present model which does not seem to be currently
 justified.  While cognizant of the above deficiencies,
 we wish to believe that they are outweighted by the model merits.

  \newpage 
  
  \begin{table}
  \caption{
  $\pi$d scattering length 
  obtained from the static model and from a Faddeev calculation
  in the zero-range model for different $b_{0}$ and $b_{1}$.
  For the NN forces we used PEST and Yamaguchi potentials,
  the results for the latter case are presented here in brackets.
  All entries are in $10^{-2}/m_{\pi}$ units.}
  \label{table1}
  \vspace*{2mm}
  \begin{tabular}{cccccc}
  \multicolumn{6}{c}{}\\
 &           &                &               & $b_{1}$      &       \\
 &           &                &               &              &       \\
  \hline
 &           &                &               &              &       \\
 & $b_{0}$   &  model         & -9.47         & -9.05        &  -8.63\\
 &           &                &               &              &       \\
 \hline 
 &       & 2-nd order          & -4.22 (-4.87) &-3.97 (-4.57) &-3.74 (-4.28)\\  
 &    &static \Ref{a10}        & -3.89 (-4.21) &-3.69 (-3.98) &-3.49 (-3.77)\\
 &-0.65&static \Ref{a26}       & -3.44 (-3.77) &-3.29 (-3.58) &-3.10 (-3.39)\\
 & & Faddeev                   & -3.97 (-4.27) &-3.76 (-4.04) &-3.55 (-3.81)\\  
 &  &{\em ditto} with $\Delta$ & -3.59 (-3.97) &-3.37 (-3.73) &-3.16 (-3.50)\\  
 \hline 
 &       & 2-nd order          & -3.30 (-3.96) &-3.06 (-3.66) &-2.82 (-3.37)\\ 
 &    &static \Ref{a10}        & -2.99 (-3.32) &-2.78 (-3.09) &-2.58 (-2.87)\\
 &-0.22&static \Ref{a26}       & -2.53 (-2.87) &-2.36 (-2.68) &-2.19 (-2.49)\\
 &       & Faddeev             & -3.07 (-3.37) &-2.85 (-3.14) &-2.65 (-2.92)\\
 &  &{\em ditto} with $\Delta$ & -2.68 (-3.08) &-2.46 (-2.85) &-2.25 (-2.62)\\
 \hline 
 &       & 2-nd order          & -2.38 (-3.04) &-2.14 (-2.74) &-1.90 (-2.45)\\
 & &static \Ref{a10}           & -2.08 (-2.42) &-1.87 (-2.19) &-1.68 (-1.97)\\
 & 0.21 &static \Ref{a26}      & -2.62 (-1.97) &-1.45 (-1.77) &-1.28 (-1.59)\\
 &       & Faddeev             & -2.16 (-2.47) &-1.95 (-2.24) &-1.74 (-2.02)\\
 &  &{\em ditto} with $\Delta$ & -1.76 (-2.20) &-1.54 (-1.96) &-1.34 (-1.73)\\
  \multicolumn{6}{c}{}\\
  \end{tabular}
  \end{table}

  \begin{table}
  \caption{ 
  The expectation values of r, 1/r 
  and the values of $\pi$d scattering length
  calculated for different NN wavefunctions.
  For $\pi$N scattering lengths we have adopted their central values, i.e.
   $b_{0}=-0.22$ and $b_{1}=-9.05$. 
   All scattering lengths are given in $10^{-2}/m_{\pi}$ units.
   }
  \label{table2}
  \vspace*{2mm}
  \begin{tabular}{cccccc}
  \multicolumn{6}{c}{}\\
 & & &    NN wavefunction & & \\ 
 &&&&&\\
 \hline 
 &        &   Hulthen   & PEST      & Paris    &     Bonn      \\
 \hline 
 &$\bra r \ket fm$ & 3.1345 & 3.2309 & 3.2685 &  3.2536 \\
 &$\bra 1/r \ket fm^{-1}$ & 0.55501 & 0.45507 & 0.44864 &  0.46314 \\
 & 2-nd order $a_{\pi d}$  & -3.66 & -3.06  & -3.04   &  -3.13     \\
 & static $a_{\pi d}$ &   -3.09     & -2.78     & -2.78    & -2.82      \\
  \multicolumn{6}{c}{}\\
  \end{tabular}
  \end{table}

  \begin{table}
  \caption{ 
  $\pi$d scattering lengths calculated from  
   consecutive iterations of the Faddeev equations.
  All entries are in $10^{-2}/m_{\pi}$ units.   
  }
  \label{table3}
  \vspace*{2mm}
  \begin{tabular}{cccccc}
  \multicolumn{6}{c}{}\\
  &      &   PEST&PEST & Yamaguchi&Yamaguchi\\ 
 & order & no $\Delta$ & with $\Delta$ & no $\Delta$ & with $\Delta $\\
 &&&&&\\
  \hline
 & 1  & -1.66 & -1.21 & -1.70 & -1.23\\
 & 2  & -2.98 & -2.66 & -3.42 & -3.30\\
 & 3  & -2.89 & -2.44 & -3.20 & -2.77\\
 & 4  & -2.85 & -2.48 & -3.11 & -2.91\\
 & 5  & -2.85 & -2.45 & -3.14 & -2.82\\
 & 6  &       & -2.46 & -3.15 & -2.87\\
 & 7  &       & -2.46 & -3.14 & -2.84\\
 & 8  &       &       & -3.14 & -2.86\\
 & 9  &       &       &       & -2.85\\
 & 10 &       &       &       & -2.85\\
  \multicolumn{6}{c}{}\\
  \end{tabular}
  \end{table}

  \begin{table}
  \caption{ $\pi$N scattering lengths inferred from pionic
  hydrogen data ($B_{2I}$ are slope parameters defined in the text). 
  }
  \label{table4}
  \vspace*{2mm}
  \begin{tabular}{cccccc}
  \multicolumn{6}{c}{}\\
 & $\beta$     & a $_{1}$ & a$_{3}$ &  B$_{1}$ & B$_{3}$  \\
 & [fm$^{-1}$]&[m$_{\pi}^{-1}$]&[10$^{-1}\,$m$_{\pi}^{-1}$]&
 [10$^{-2}\,$m$_{\pi}^{-3}]$& [10$^{-2}\,$m$_{\pi}^{-3}$]\\
  \hline
&  2.0&      0.1767$\pm$      0.0046&      -0.9377$\pm$      0.0852
 &     -6.63      &      1.96\\
&  3.0&      0.1760$\pm$      0.0046&      -0.9306$\pm$      0.0846
 &     -3.60      &      0.81\\
&  4.0&      0.1757$\pm$      0.0046&      -0.9263$\pm$      0.0841
 &     -2.46      &      0.43\\
&  5.0&      0.1756$\pm$      0.0046&      -0.9228$\pm$      0.0837
 &     -1.90      &      0.27\\
&  6.0&      0.1756$\pm$      0.0046&      -0.9197$\pm$      0.0834
 &     -1.57      &      0.18\\
&  7.0&      0.1756$\pm$      0.0046&      -0.9167$\pm$      0.0830
 &     -1.37      &      0.14\\
&  8.0&      0.1756$\pm$      0.0046&      -0.9138$\pm$      0.0827
 &     -1.23      &      0.11\\
&  9.0&      0.1756$\pm$      0.0047&      -0.9110$\pm$      0.0823
 &     -1.12      &      0.09\\
& 10.0&      0.1757$\pm$      0.0047&      -0.9082$\pm$      0.0820
 &     -1.05      &      0.08\\
 \hline
 & Deser & 0.1760 $\pm$ 0.0046&-0.9258 $\pm$0.0857& &  \\
 \hline
 & ref. \cite{Gashi}& 0.1679 $\pm$ 0.0059 &-0.785 $\pm$ 0.034 &
 -7.24 $\pm$ 3.06 & -4.08 $\pm$ 1.46\\
  \multicolumn{6}{c}{}\\
  \end{tabular}
  \end{table}

 \newpage

 \newpage
 \begin{figure}[ht]
 \centering
 \caption{$\sin \delta$ vs energy close to the resonance  
 calculated from \Ref{b16} for $\beta=3\,fm^{-1}$.
      }
 	 \label{fig2}
  	    \vspace*{8mm}

 \includegraphics[width=0.5\textwidth,totalheight=0.5\textheight]{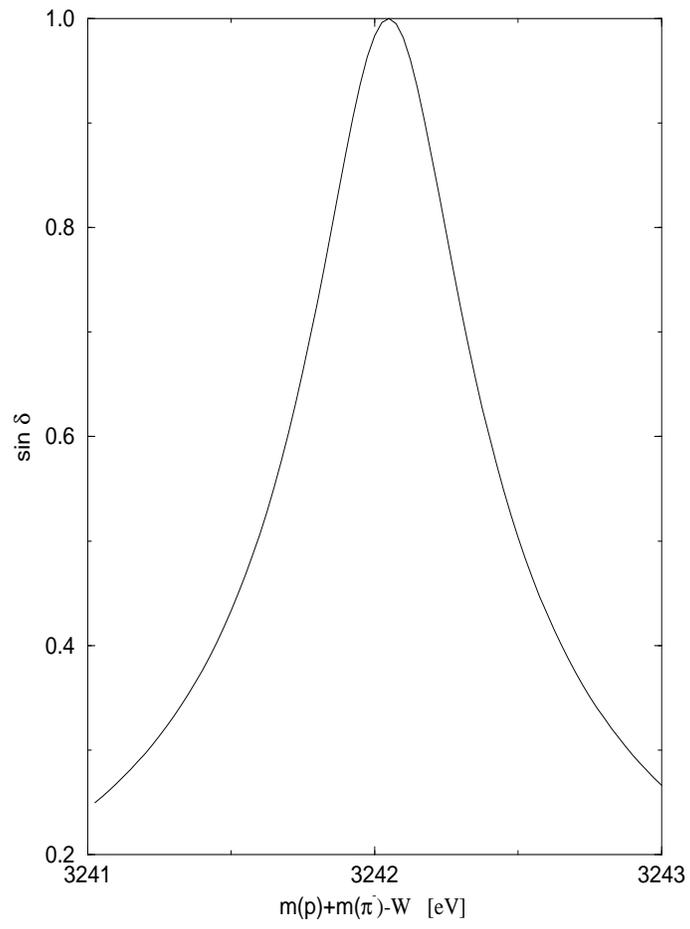}
 	       \end{figure}
 \newpage
 \begin{figure}[ht]
 \centering
 \caption{ 
  The values of isoscalar and isovector scattering lengths obtained
  by solving the bound state equation \Ref{b5}
  for $\beta$ equal, respectvely, 2.0 fm$^{-1}$, 6.0 fm$^{-1}$,
  and 10.0 fm$^{-1}$ (indicated on the plot).
  The point marked as Deser has been obtained from \Ref{b31}.
  The rectangle corresponds to the values  obtained in
  \protect\cite{Sigg}.
  }
 	 \label{fig3}
  	    \vspace*{8mm}

 \includegraphics[width=0.4\textwidth ,totalheight=0.4\textheight ]{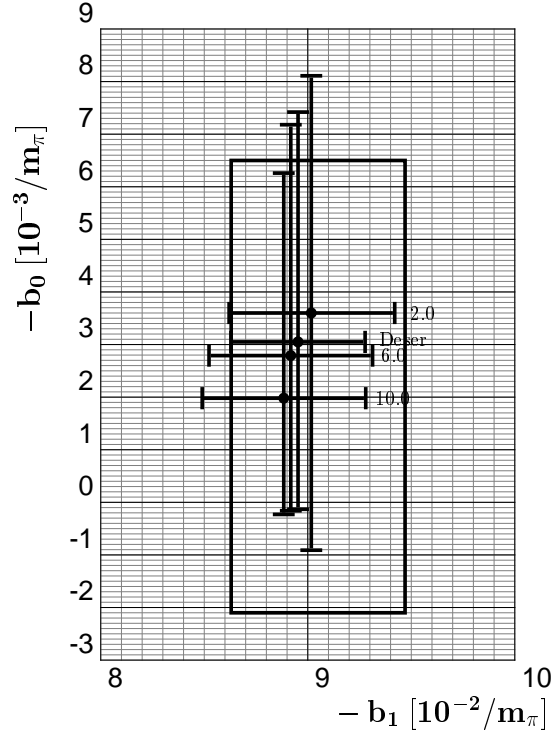}
 	       \end{figure}
 \newpage
 \begin{figure}[ht]
 \centering
 \caption{ $\pi$d scattering length vs. the inverse range parameter
    $\beta$ of the $\pi$N potential. Full (open) circles correspond 
    to a Faddeev calculation with (without) p-wave $\pi$N interaction.
      }
 	 \label{fig4}
  	    \vspace*{8mm}

 \includegraphics[width=0.4\textwidth,totalheight=0.4\textheight]{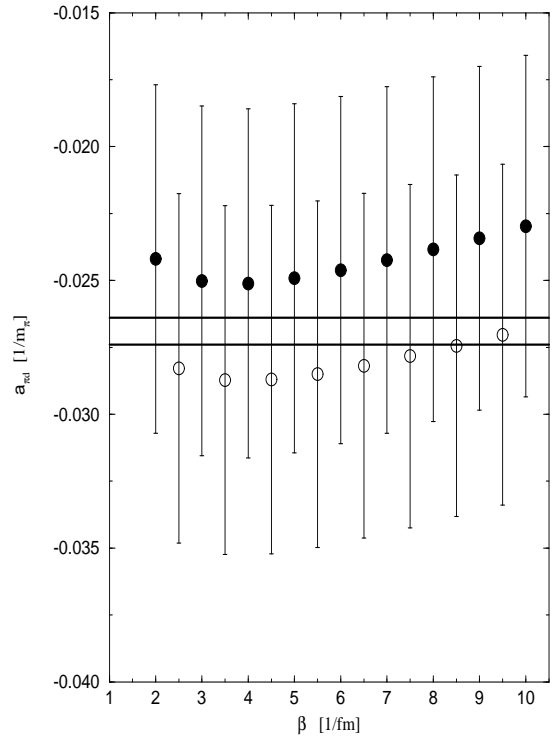}
 	       \end{figure}

  
 \end{document}